\documentclass[journal,10pt,a4paper,twoside]{IEEEtran}

\ifCLASSOPTIONdraftcls
	\newcommand{\figheight}{85mm} 
	
\else
  \newcommand{\figheight}{70mm}
  
\fi

\usepackage{ifthen}

\usepackage{amsmath}
\usepackage{amssymb}
\usepackage{amsfonts}
\usepackage{graphicx}
\usepackage{psfrag}
\usepackage{paralist}
\usepackage{times}
\usepackage{amsbsy}
\usepackage{latexsym,bm}
\usepackage{ifsym}
\usepackage{epsfig}
\usepackage{color}
\usepackage{paralist}
\usepackage{psfrag}
\usepackage{cite}
\usepackage{stfloats}

\usepackage{amsthm}
\newtheorem{thm}{Theorem}

\theoremstyle{definition}

\usepackage{flushend}

\usepackage{bm}

\newcommand{\compl}{\mathbb{C}}        
\newcommand{\real}{\mathbb{R}}         

\newcommand{\e}{{\rm e}} 
\renewcommand{\j}{\mathrm{j}} 

\newcommand{\opof}[2]{\mathop{{\rm #1}}\left\{#2\right\}}         

\newcommand{\realof}[1]{\opof{Re}{#1}}           
\newcommand{\imagof}[1]{\opof{Im}{#1}}           
\newcommand{\diagof}[1]{\opof{diag}{#1}}         
\newcommand{\expvof}[1]{\opof{\mathbb{E}}{#1}}   

\newcommand{\vecof}[1]{\opof{vec}{#1}}           

\newcommand{\expof}[1]{{\e}^{#1}}                      

\newcommand{\subsel}[1]{#1^{({\rm sel})}}

\newcommand{\sub}[1]{#1_{\rm sub}}
\newcommand{\subr}[1]{#1_{{\rm sub}_r}}
\newcommand{\subp}[1]{#1_{{\rm sub}_p}}
\newcommand{\subl}[1]{#1_{{\rm sub}_l}}
\newcommand{\ssm}[1]{{#1}_{{\rm SS}}}
\newcommand{\ssms}[1]{#1_{{\rm SS}_{\rm s}}}
\newcommand{\ssmn}[1]{#1_{{\rm SS}_{\rm n}}}
\newcommand{\ssmnH}[1]{#1_{{\rm SS}_{\rm n}}^\herm}
\newcommand{\ssmone}[1]{{#1}_{{{\rm SS}_1}}}
\newcommand{\ssmtwo}[1]{{#1}_{{{\rm SS}_2}}}
\newcommand{\ssmk}[1]{{#1}_{{{\rm SS}_k}}}
\newcommand{\ssmi}[1]{{#1}_{{{\rm SS}_i}}}
\newcommand{\ssmnought}[1]{{#1}_{{{\rm SS}_0}}}
\newcommand{\ssmtil}[1]{\tilde{#1}_{\rm SS}}

\newcommand{\ssmtils}[1]{\tilde{#1}_{{\rm SS}_{\rm s}}}
\newcommand{\ssmtiln}[1]{\tilde{#1}_{{\rm SS}_{\rm n}}}
\newcommand{\ssmtilsinv}[1]{\tilde{#1}_{{\rm SS}_{\rm s}}^{-1}}

\newcommand{\ssmnc}[1]{{#1}_{{\rm SS}}^{({\rm nc})}}

\newcommand{\ssmncH}[1]{{#1}_{{\rm SS}}^{({\rm nc})^\herm}}
\newcommand{\ssmncC}[1]{{#1}_{{\rm SS}}^{({\rm nc})^\conj}}
\newcommand{\ssmncT}[1]{{#1}_{{\rm SS}}^{({\rm nc})^\trans}}
\newcommand{\ssmncs}[1]{{#1}_{{\rm SS}_{\rm s}}^{({\rm nc})}}
\newcommand{\ssmncn}[1]{{#1}_{{\rm SS}_{\rm n}}^{({\rm nc})}}

\newcommand{\ssmncsT}[1]{{#1}_{{\rm SS}_{\rm s}}^{({\rm nc})^\trans}}
\newcommand{\ssmncnH}[1]{{#1}_{{\rm SS}_{\rm n}}^{({\rm nc})^\herm}}

\newcommand{\ssmncsinv}[1]{{#1}_{{\rm SS}_{\rm s}}^{({\rm nc})^{-1}}}

\newcommand{\sigman}{{\sigma}_{{\rm n}}}

\newcommand{\rd}[1]{#1^{({r})}}

\newcommand{\rdT}[1]{#1^{({r})^\trans}}
\newcommand{\rdH}[1]{#1^{({r})^\herm}}
\newcommand{\rdC}[1]{#1^{({r})^\conj}}
\newcommand{\rdtil}[1]{\tilde{#1}^{({r})}}
\newcommand{\rdtilH}[1]{\tilde{#1}^{({r})^\herm}}
\newcommand{\rdtilT}[1]{\tilde{#1}^{({r})^\trans}}

\newcommand{\nc}[1]{#1^{({\rm nc})}}
\newcommand{\ncr}[1]{#1^{({\rm nc})(r)}}
\newcommand{\ncrT}[1]{#1^{({\rm nc})(r)^\trans}}

\newcommand{\ncrH}[1]{#1^{({\rm nc})(r)^\herm}}
\newcommand{\ncrtil}[1]{\tilde{#1}^{({\rm nc})(r)}}

\newcommand{\ncT}[1]{#1^{({\rm nc})^\trans}}
\newcommand{\ncC}[1]{#1^{({\rm nc})^\conj}}
\newcommand{\ncH}[1]{#1^{({\rm nc})^\herm}}

\newcommand{\normof}[2]{\left\|#1\right\|_{#2}}
\newcommand{\twonorm}[1]{\normof{#1}{2}}                  

\newcommand{\inv}{{-1}}          
\newcommand{\pinv}{+}          
\newcommand{\conj}{*}          
\newcommand{\trans}{{\rm T}}   
\newcommand{\herm}{{\rm H}}    

\usepackage{color}
\newcommand{\red}[1]{{#1}}

\title{Performance Analysis of Multi-Dimensional ESPRIT-Type Algorithms for Arbitrary and Strictly Non-Circular Sources with Spatial Smoothing}
\author{Jens Steinwandt$^*$,~\IEEEmembership{Student Member,~IEEE},
		    Florian Roemer,~\IEEEmembership{Senior Member,~IEEE},\\ 
		    Martin Haardt,~\IEEEmembership{Senior Member,~IEEE}, and 
		    Giovanni Del Galdo,~\IEEEmembership{Member,~IEEE}

\thanks{
Minor parts of this paper have been published at the 2014 {\em IEEE Int. Conf. on Acoustics, Speech, 
and Signal Proc. (ICASSP)} \cite{steinwandt2014spasmoo}.
}
\thanks{
The authors J.~Steinwandt, F.~Roemer, M.~Haardt, and G.~Del Galdo are with Ilmenau University of Technology,
P.O.~Box 100565, D-98684 Ilmenau, Germany,
e-mail: \{jens.steinwandt, florian.roemer, martin.haardt, giovanni.delgaldo\}@tu-ilmenau.de,
phone: +49 (3677) 69-2613, web: http://www.tu-ilmenau.de/crl and http://www.tu-ilmenau.de/dvt.

F.~Roemer and G.~Del Galdo are also with the Fraunhofer Institute for Integrated Circuits IIS.
}
\thanks{$*$ corresponding author}
}

\begin{document}

\maketitle
\linespread{1}

\vspace{-1cm}

%
\begin{abstract}
Spatial smoothing is a widely used preprocessing scheme to improve the performance of high-resolution 
parameter estimation algorithms in case of coherent signals or if only a small number of snapshots 
is available. In this paper, we present a first-order performance analysis of the spatially smoothed 
versions of $R$-D Standard ESPRIT and $R$-D Unitary ESPRIT for sources with arbitrary signal 
constellations as well as $R$-D NC Standard ESPRIT and $R$-D NC Unitary ESPRIT for strictly 
second-order (SO) non-circular (NC) sources. 
The derived expressions are asymptotic in the effective signal-to-noise ratio (SNR), i.e., the 
approximations become exact for either high SNRs or a large sample size. Moreover, no assumptions on 
the noise statistics are required apart from a zero-mean and finite SO moments. We show that both 
$R$-D NC ESPRIT-type algorithms with spatial smoothing perform asymptotically identical in the high 
effective SNR regime. Generally, the performance of spatial smoothing based algorithms depends on the 
number of subarrays, which is a design parameter and needs to be chosen beforehand. In order to gain 
more insights into the optimal choice of the number of subarrays, we simplify the derived analytical 
$R$-D mean square error (MSE) expressions for the special case of a single source. The obtained MSE 
expression explicitly depends on the number of subarrays in each dimension, which allows us to analytically 
find the optimal number of subarrays for spatial smoothing. Based on this result, we additionally derive 
the maximum asymptotic gain from spatial smoothing and explicitly compute the asymptotic efficiency for 
this special case. All the analytical results are verified by simulations.
%
%
\end{abstract}
%

\begin{IEEEkeywords}
Spatial smoothing, ESPRIT, non-circular sources, performance analysis, DOA estimation.
\end{IEEEkeywords}

\IEEEpeerreviewmaketitle


\section{Introduction} 
\label{sec:intro}
\IEEEPARstart{T}{he} problem of high resolution parameter estimation from multi-dimensional ($R$-D) signals with 
$R \geq 1$ has long been a fundamental research area in the field of array signal processing. Such a task, e.g., 
estimating the directions of arrival, directions of departures, frequencies, Doppler shifts, etc. arises in a \red{wide 
range of applications including radar \cite{nion2010mimoradar}, sonar \cite{cox1989sonar}, channel sounding 
\cite{haardt2004channel,liu2004channel}, and wireless communications \cite{pesavento2004mimo}.} 
$R$-D ESPRIT-type parameter estimation algorithms \cite{haardt1998schur} have attracted considerable attention due 
to their fully algebraic estimates and their low complexity. Hence, their analytical performance assessment has also 
been of great research interest. Two fundamental performance analysis concepts for 1-D parameter estimation have been 
established in \cite{rao1989perf} and \cite{vaccaro1993perf}. While \cite{rao1989perf} relies on the eigenvector 
distribution of the sample covariance matrix and is only asymptotic in the sample size $N$, the framework in 
\cite{vaccaro1993perf} provides an explicit first-order approximation of the parameter estimation error based on the 
superposition of the signal component by a small noise perturbation. The latter is asymptotic in the effective 
signal-to-noise ratio (SNR), i.e., the results become accurate for either high SNRs or a large sample size. Therefore, 
\cite{vaccaro1993perf} is more general than \cite{rao1989perf} as it is even valid for $N=1$ if the SNR is sufficiently 
high. In \cite{roemer2012perf,roemer2014perfana}, this performance analysis framework was extended to $R$-D parameter 
estimation, where no assumptions on the noise statistics apart from a zero mean and finite second-order (SO) moments 
are required for the analytical mean square error (MSE) expressions.

Many authors have 
shown that taking advantage of existing properties of the observed signals such as their 
strictly SO non-circular (NC) structure \cite{schreier2010noncirc} helps to improve the performance of conventional 
parameter estimation algorithms. Examples of such NC signals include BPSK, PAM, and ASK-modulated signals. 
\red{They are of practical relevance 
in wireless communications, cognitive radio, GNSS satellite systems etc., when strictly non-circular sources are known to be 
present, or in radar, tracking, channel sounding, etc., where the transmit signals can be designed as strictly non-circular.}
Recently, a 
number of improved subspace-based parameter estimation schemes, e.g., \red{NC MUSIC \cite{abeida2006perf,ferreol2010nc2qmusic,ferreol2014insights},} NC Root-MUSIC 
\cite{charge2001ncroot}, NC Standard ESPRIT \cite{zoubir2003ncesprit}, and NC Unitary ESPRIT \cite{haardt2004ncunit,steinwandt2014tsp} 
have been developed. It has been demonstrated that exploiting the prior knowledge on the signals' strict non-circularity 
significantly improves the estimation accuracy and doubles the number of identifiable sources \cite{haardt2004ncunit}. 
The analytical performance of the MUSIC and ESPRIT-based NC algorithms has been investigated in \cite{abeida2006perf, 
abeida2008perf,steinwandt2014tsp}. For the special case of a single source, it was shown in \cite{roemer2012perf} along 
with \cite{steinwandt2014tsp} that neither forward-backward averaging (FBA) nor NC preprocessing in combination with 
ESPRIT-type algorithms improve the asymptotic MSE. \red{The more general case of coexisting circular and strictly 
non-circular signals has been considered in \cite{gao2008coex,steinwandt2015coex,steinwandt2016coexperf}.}

The aforementioned NC and \red{conventional} methods are known to yield a high resolution even in the case of correlated sources. 
However, they fail when more than two signals\footnote{Two coherent signals can be separated by forward-backward 
averaging (FBA) if the array phase reference is not located at the array centroid \cite{haardt1997phd}.} are coherent 
(fully correlated) or if $N=1$, as both render the signal covariance matrix rank-deficient. \red{In practice, coherent signals 
often occur in a multipath environment \cite{tse2005mimo} and the single snapshot case is often encountered in, e.g., 
channel sounding \cite{liu2004channel}, co-prime arrays \cite{pal2010nested}, tracking \cite{comon1990tracking}. }
Assuming a uniform array geometry, preprocessing via spatial smoothing 
\cite{evans1982spasmoo,shan1985spasmoo,pillai1989spasmo} can be applied to estimate the parameters of coherent signals. 
Spatial smoothing decorrelates coherent signals by averaging the data received by a number of subarrays $L$. As the resulting estimation 
error depends on $L$, this is a design parameter that can be optimized to achieve the best estimation accuracy. Several 
performance analyses of parameter estimation schemes using spatial smoothing based on the framework \cite{rao1989perf}, 
which is, however, only asymptotic in $N$, have been presented in \cite{pillai1989spasmoo,rao1990spasmoo,rao1993spasmoo, 
hua1990spasmoo, hari1999spasmoo, lemma2003spasmoo, weiss1993spasmoointerpol}. While \cite{pillai1989spasmoo,rao1990spasmoo, 
rao1993spasmoo} consider spatially smoothed MUSIC-type algorithms, the references \cite{hua1990spasmoo,hari1999spasmoo, 
lemma2003spasmoo} study ESPRIT-type algorithms. In \cite{weiss1993spasmoointerpol}, a performance analysis for an 
interpolated spatial smoothing algorithm for non-uniform linear arrays was proposed. The special case of spatial smoothing 
for a single source was considered in \cite{rao1990spasmoo, rao1993spasmoo}, and in \cite{hua1990spasmoo} for harmonic 
retrieval. It was observed that in this case a gain from spatial smoothing can be achieved. However, these existing 
performance analysis results only concern the 1-D parameter estimation. Analytical expressions for $R$-D parameter 
estimation algorithms such as $R$-D Standard ESPRIT and $R$-D Unitary ESPRIT with spatial smoothing as well as 
their recently proposed NC-versions $R$-D NC Standard ESPRIT and $R$-D NC Unitary ESPRIT with spatial smoothing 
have not been reported in the literature.

Therefore, in this paper, we present a first-order performance analysis for the spatially smoothed versions of $R$-D 
Standard ESPRIT and $R$-D Unitary ESPRIT as well as $R$-D NC Standard ESPRIT and $R$-D NC Unitary ESPRIT 
based on the more general framework in \cite{vaccaro1993perf}, which is asymptotic in the high effective SNR. We 
assume a uniform $R$-D array geometry and use least squares (LS) to solve the shift invariance equations. \red{However, 
as LS and total least squares (TLS) have been shown to perform asymptotically identical \cite{rao1989perf}, the results 
obtained for LS are also valid for TLS.} The derived 
closed-form MSE expressions are explicit in the noise realizations such that apart from a zero mean and finite SO 
moments, no further assumptions on the noise statistics are required. We show that due to the NC preprocessing both 
$R$-D NC ESPRIT-type algorithms with spatial smoothing perform identical in the high effective SNR. Further insights 
into the dependence of the MSE expressions on the physical parameters are provided by the case study of a single source. 
For this case, we first show that $R$-D spatial smoothing improves the estimation accuracy and that all the considered 
spatial smoothing based $R$-D ESPRIT-type algorithms provide the same MSE result, i.e., asymptotically, no additional 
gain is obtained from FBA and NC preprocessing. Based on these results, we analytically find the optimal number of 
subarrays $L$ that minimizes the MSE in each of the $R$ dimensions, which extends the 1-D results in \cite{pillai1989spasmoo, 
rao1990spasmoo, rao1993spasmoo, hua1990spasmoo, hari1999spasmoo, lemma2003spasmoo, weiss1993spasmoointerpol}. This enables us to compute 
the maximum asymptotic $R$-D spatial smoothing gain for a single source in closed-form. Additionally, we analytically 
compute the asymptotic efficiency of the spatial smoothing based algorithms for $R=1$.

This paper is organized as follows: The $R$-D data model and the preprocessing for NC sources are introduced in Section 
\ref{sec:data}. Section \ref{sec:spasmoo} reviews $R$-D spatial smoothing for ESPRIT-type and NC ESPRIT-type algorithms. 
Their performance analysis is presented in Section \ref{sec:perf} before the special case of a single source is analyzed 
in Section \ref{sec:single}. Section \ref{sec:simulations} illustrates the numerical results, and concluding remarks are 
drawn in Section \ref{sec:conclusions}.

\textit{Notation:} We use lower-case bold-face letters for column vectors and upper-case bold-face letters for matrices. 
The superscripts $^\trans$, $^\conj$, $^\herm$, $^{-1}$, $^+$ denote the transposition, complex conjugation, conjugate 
transposition, matrix inversion, and the Moore-Penrose pseudo inverse, respectively. The Kronecker product is denoted as 
$\otimes$ and the Khatri-Rao product (column-wise Kronecker product) as $\diamond$. The operator $\vecof{\bm A}$ stacks the 
columns of the matrix into a large column vector, 
$\mathrm{diag}\{\bm a\}$ returns a diagonal matrix with the elements of $\bm a$ placed on its diagonal, and 
$\mathrm{blkdiag}\{\cdot\}$ creates a block diagonal matrix. The operator $\mathcal O\{\cdot\}$ denotes the highest order 
with respect to a parameter. The matrix $\bm \Pi_M$ is the $M \times M$ exchange matrix with ones on its anti-diagonal and 
zeros elsewhere and $\underline{\bm 1}$ denotes the vector of ones. Moreover, $\realof{\cdot}$ and $\imagof{\cdot}$ 
extract the real and imaginary part of a complex number and $\mathrm{arg}\{\cdot\}$ extracts its phase. Also, $\twonorm{\bm x}$ 
represents the 2-norm of the vector $\bm x$, and $\expvof{\cdot}$ stands for the statistical expectation. Furthermore, we use the 
short hand notation 
\begin{align}
\sum_{\underline{\bm \ell} = \underline{\bm 1}}^{\underline{\bm L}} x_{\underline{\bm \ell}} = \sum_{\ell_1 = 1}^{L_1} \sum_{\ell_2 = 1}^{L_2} \cdots 
\sum_{\ell_R = 1}^{L_R} x_{\ell_1,\ldots,\ell_R},
\end{align}
where $\underline{\bm \ell} = [\ell_1,\ldots,\ell_r,\ldots,\ell_R]$ and $\underline{\bm L} = [L_1,\ldots,L_r,\ldots,L_R]$ with 
$\ell_r = 1,\ldots,L_r,~ r = 1,\ldots,R$. 
\section{Data Model}
\label{sec:data}
In this section, we introduce the $R$-D data model for arbitrary signals followed by the NC data model 
for strictly non-circular signals.
\subsection{Data Model for Arbitrary Signals}
Suppose the measurement data is represented by $N$ subsequent observations of a noise-corrupted 
superposition of $d$ undamped exponentials sampled on a separable uniform $R$-D grid of size 
$M_1\times\ldots\times M_R$ \cite{haardt1998schur}. The $t_n$-th time snapshot of the $R$-D 
measurements can be modeled as

\vspace{-1em}
\small
\begin{align}
x_{m_{1},\ldots,m_{R}}(t_n) \! = \! \sum_{i=1}^d s_i(t_n) \! \prod_{r=1}^R \expof{\j (m_r - 1)\rd{\mu}_i} \!\! + n_{m_{1},\ldots,m_{R}}(t_n),
\label{rdmodel}
\end{align}
\normalsize
%
where $m_r = 1,\ldots,M_r$, $n=1,\ldots,N$, and $s_i(t_n)$ represents the complex amplitude of the 
$i$-th undamped exponential at the time instant $t_n$. Furthermore, $\rd{\mu}_i$ is the $i$-th spatial 
frequency in the $r$-th mode, $i=1,\ldots,d$, $r=1,\ldots,R$, and $n_{m_{1},\ldots,m_{R}}(t_n)$ 
denotes the zero-mean additive noise component. In the context of array signal processing, each 
of the $R$-D exponentials represents a narrow-band planar wavefront from a stationary far-field 
source and the complex amplitudes $s_i(t_n)$ describe the source symbols. The goal is to estimate 
the $R {\cdot} d$ spatial frequencies $\bm \mu_i = [\mu^{(1)}_i,\ldots,\mu^{(R)}_i]^\trans,~\forall i$, 
from \eqref{rdmodel}. We assume that $d$ is known or has been estimated beforehand.

In order to obtain a more compact formulation of \eqref{rdmodel}, we form the measurement matrix 
$\bm X \in\compl^{M \times N}$ with $M = \prod_{r=1}^R M_r$ by stacking the $R$ spatial dimensions 
and aligning the $N$ time snapshots as the columns. This way, $\bm X$ can be 
modeled as 
\begin{align}
\bm X = \bm A \bm S + \bm N ~ \in\compl^{M \times N},
\label{model}
\end{align}
where $\bm S \in\compl^{d \times N}$ represents the source symbol matrix, $\bm N \in\compl^{M \times N}$ 
contains the noise samples, and $\bm A = [\bm a(\bm \mu_1), \ldots, \bm a(\bm \mu_d)] \in\compl^{M \times d}$ 
is the array steering matrix. The latter consists of the array steering vectors $\bm a(\bm \mu_i)$ 
corresponding to the $i$-th spatial frequency, which are given by
\begin{align}
\bm a(\bm \mu_i) = \bm a^{(1)} \left(\mu_i^{(1)}\right) \otimes \cdots \otimes \bm a^{(R)} \left(\mu_i^{(R)} \right)~ \in\compl^{M \times 1},
\label{steer}
\end{align}
where $\rd{\bm a}(\rd{\mu}_i)\in\compl^{M_r \times 1}$ is the array steering vector in the $r$-th mode. 
Alternatively, $\bm A$ can be expressed as
\begin{align}
\bm A = \bm A^{(1)} \diamond \bm A^{(2)} \diamond \cdots \diamond \bm A^{(R)},
\label{steermat}
\end{align}
where $\rd{\bm A} = [\rd{\bm a}(\rd{\mu_1}),\ldots,\rd{\bm a}(\rd{\mu_d})] 
\in\mathbb C^{M_r\times d}$ represents the array steering matrix in the $r$-th mode.
For an arbitrary phase reference along the $r$-th mode, $\rd{\bm A}$ can be decomposed as \cite{steinwandt2015rdnccrb}
%
$\rd{\bm A} = \rd{\bar{\bm A}} \rd{\bm \Delta}$,
%
where $\rd{\bar{\bm A}} = [\rd{\bar{\bm a}}(\rd{\mu}_1),\cdots,\rd{\bar{\bm a}}(\rd{\mu}_d)] \in\compl^{M_r \times d}$ 
satisfies $\rd{\bar{\bm A}} = \bm \Pi_{M_r} \rdC{\bar{\bm A}}$ and contains the steering vectors $\rd{\bar{\bm a}}
(\rd{\mu}_i),~i=1,\ldots,d,$ whose phase reference is located at the centroid of the $r$-th mode, i.e.,
\begin{align} 
\rd{\bar{\bm a}}(\rd{\mu}_i) = \begin{bmatrix} \e^{-\j\frac{(M_r-1)}{2}\rd{\mu}_i} & \cdots & \e^{\j\frac{(M_r-1)}{2} \rd{\mu}_i} 
\end{bmatrix}. 
\label{asteer_phasecen}
\end{align}
Furthermore, the diagonal matrix $\rd{\bm \Delta} = \mathrm{diag}\big\{\e^{\j \rd{\delta} \rd{\mu}_i} \big\}_{i=1}^d$ 
defines the shifts of the phase reference $\rd{\delta} \in [\frac{-(M_r-1)}{2},\frac{(M_r-1)}{2}]$ for each $\rd{\mu}_i$. 
If the actual phase reference is at the array centroid of the $r$-th mode, we have $\rd{\delta} = 0$, 
$\rd{\bm \Delta} = \bm I_d$, and consequently $\rd{\bm A} = \rd{\bar{\bm A}}$.
Thus, we can rewrite $\bm A$ in \eqref{steermat} as \cite{steinwandt2015rdnccrb}
%
$\bm A = \bar{\bm A} \bm \Delta$,
%
where $\bar{\bm A} = \bar{\bm A}^{(1)} \diamond \bar{\bm A}^{(2)} \diamond \cdots \diamond \bar{\bm A}^{(R)} 
\in\compl^{M \times d}$ and $\bm \Delta = \bm \Delta^{(1)} \cdot \bm \Delta^{(2)} \cdot \ldots \cdot
\bm \Delta^{(R)}\in\compl^{d \times d}$. Again, if $\rd{\delta} = 0~\forall r$, we have $\bm A = \bar{\bm A}$.
Using these relations, we obtain the model 
\begin{align}
\bm X = \bar{\bm A} \bm \Delta \bm S + \bm N = \bar{\bm A} \bar{\bm S} + \bm N ~ \in\compl^{M \times N}.
\label{model2}
\end{align}

Due to the assumption that the $R$-D sampling grid is uniform, the array steering matrix $\bm A$ satisfies 
the shift invariance equations given by 
\begin{align}
\rdtil{\bm J}_1 \bar{\bm A} ~\rd{\bm \Phi} = \rdtil{\bm J}_2 \bar{\bm A}, \quad r = 1,\ldots,R,
\label{shift}
\end{align}
where $\rdtil{\bm J}_1$ and $\rdtil{\bm J}_2 \in\real^{\frac{M}{M_r}(M_r-1)\times M}$ are the 
effective $R$-D selection matrices, which select $M_r-1$ elements (maximum overlap) for the first and 
the second subarray in the $r$-th mode, respectively. 
They are compactly defined as $\rdtil{\bm J}_k=\bm I_{\prod_{l=1}^{r-1} M_l} \otimes \rd{\bm J}_k \otimes 
\bm I_{\prod_{l=r+1}^{R} M_l}$ for $k=1,2$, where $\rd{\bm J}_k \in\real^{(M_r-1)\times M_r}$ are 
the $r$-mode selection matrices for the first and second subarray \cite{haardt1998schur}. The diagonal matrix $\rd{\bm \Phi}  =\mathrm{diag}\{[\expof{\j\rd{\mu}_1},\ldots,\expof{\j\rd{\mu}_d}]\}\in\mathbb C^{d\times d}$ contains 
the spatial frequencies in the $r$-th mode to be estimated. 

%
%
\subsection{Preprocessing for Strictly Non-Circular Signals}
\red{A zero-mean complex random variable $Z = X + \j Y$ is said to be SO non-circular if $\expvof{Z^2} \neq 0$ holds, which 
implies that its real and its imaginary part are correlated. The degree of non-circularity is usually defined by the 
non-circularity coefficient \cite{schreier2010noncirc}
\begin{align}
\kappa = \frac{\expvof{Z^2}}{\expvof{|Z|^2}} = |\kappa| \, \e^{\j\psi}, \quad 0 \leq |\kappa| \leq 1.
\end{align}
Random variables that satisfy $|\kappa|=0$ or $0 < |\kappa| < 1$ are called circularly symmetric or weak-sense SO non-circular, 
respectively. The case $|\kappa|=1$ describes a strictly SO non-circular (also referred to as rectilinear) random variable. 
The latter, which is considered in this work, implies a linear dependence between the real and the imaginary part of $Z$. Thus, 
$Z$ can be represented as a real-valued random variable $W$ which is rotated by a deterministic complex phase term 
$\e^{\j \varphi}$, i.e., $Z = W \, \e^{\j \varphi }$.

In a communication system, the case of strictly SO no-circular signals presumes that the sources transmit real-valued constellations 
(BPSK, ASK, Offset-QPSK after a derotation, etc.) whose symbol amplitudes in the complex plane at the receiver lie on lines with different phase rotations 
as the sources may have different transmission delays.} Therefore, the symbol matrix $\bm S$ in 
\eqref{model} can be decomposed as \cite{haardt2004ncunit}
%
$\bm S = \bm \Psi \bm S_0$,
%
where $\bm S_0 \in\real^{d \times N}$ is a real-valued symbol matrix and $\bm \Psi = \mathrm{diag}
\{\e^{\j\varphi_i}\}_{i=1}^d$ contains stationary complex phase shifts on its diagonal that can be 
different for each source. 
Then, $\bar{\bm S}$ in \eqref{model2} is given by 
%
$\bar{\bm S} = \bm \Delta \bm \Psi \bm S_0 = \bm \Xi \bm S_0$,
%
where we have defined $\bm \Xi = \bm \Delta \bm \Psi = \diagof{\e^{\j (\varphi_i + \delta_i )}}_{i=1}^d$ 
with $\delta_i = \sum_{r=1}^R \rd{\delta} \rd{\mu}_i$. 

In order to take advantage of the strict non-circularity of the signals, we apply a preprocessing 
scheme to \eqref{model} and define the augmented measurement matrix $\nc{\bm X} \in\compl^{2M 
\times N}$ as \cite{haardt2004ncunit} 
\red{
\begin{align}
\nc{\bm X} & = \begin{bmatrix} \bm X \\ \bm \Pi_M \bm X^\conj \end{bmatrix} 
= \begin{bmatrix} \bar{\bm A} \\ \bar{\bm A} \bm \Xi^\conj \bm \Xi^\conj \end{bmatrix} \bar{\bm S} 
+ \begin{bmatrix} \bm N \\ \bm \Pi_M \bm N^* \end{bmatrix} \notag\\ 
& = \nc{\bar{\bm A}} \bar{\bm S} + \nc{\bm N}, 
\label{ncmodel}
\end{align}
where $\bm \Pi_M$ is the $M \times M$ exchange matrix with ones on its antidiagonal and zeros 
elsewhere and we have used the property $\bm \Pi_M \bar{\bm A}^\conj = \bar{\bm A}$.} Moreover, $\nc{\bar{\bm A}} \in\compl^{2M \times d}$ and $\nc{\bm N} \in\compl^{2M \times N}$ 
are the augmented array steering matrix and the augmented noise matrix, respectively. 

It was shown in \cite{steinwandt2014tsp} that if the array steering matrix $\bar{\bm A}$ is shift-invariant 
\eqref{shift}, 
then $\nc{\bar{\bm A}}$ is also shift-invariant and satisfies
\begin{align}
\ncrtil{\bm J}_1 \nc{\bar{\bm A}} \rd{\bm \Phi} = \ncrtil{\bm J}_2 \nc{\bar{\bm A}}, \quad r = 1,\ldots,R,
\label{shiftnc}
\end{align}
where  
%
$\ncrtil{\bm J}_k = \bm I_{\prod_{l=1}^{r-1} M_l} \otimes \ncr{\bm J}_k \otimes \bm I_{\prod_{l=r+1}^{R} M_l}$ 
and $\ncr{\bm J}_k = \bm I^{}_2 \otimes \rd{\bm J}_k,~k=1,2.$
%
Note that the extended dimensions of $\nc{\bar{\bm A}}$ can be interpreted as a virtual doubling of 
the number of sensors, 
which leads to a lower estimation error and doubles the number of 
resolvable sources \cite{haardt2004ncunit}. 
%
%
%
\section{$R$-D Spatial Smoothing}
\label{sec:spasmoo}
In this section, we first apply $R$-D spatial smoothing to the data model for arbitrary signals 
in \eqref{model} before considering the strictly non-circular data model in \eqref{ncmodel}.  
\subsection{$R$-D Spatial Smoothing for Signals with Arbitrary Signal Constellations}
\label{sec:circ}
In the case of coherent signals (fully correlated), or for a single snapshot $N=1$, the 
symbol matrix $\bar{\bm S}$ becomes row rank deficient, i.e., $\mathrm{rank}\{\bar{\bm S}\}< d$. If only 
two signals are coherent, forward-backward averaging (FBA) \cite{pillai1989spasmo} can separate these signals 
if the corresponding diagonal elements of $\bm \Delta$ are distinct \cite{haardt1997phd}, i.e., the phase reference is not at the array 
centroid. For more than two coherent signals, however, the conventional subspace-based parameter estimators fail to estimate 
the directions of the coherent signals. In case of a uniform array geometry, spatial smoothing preprocessing 
can be applied to restore the full row rank $d$ of $\bar{\bm S}$ albeit reducing the effective array aperture. 

In order to perform $R$-D spatial smoothing, we apply 1-D spatial smoothing to each of the $R$ 
dimensions independently \cite{haardt1998schur}. To this end, the $M_r$ uniform sampling grid points in the $r$-th dimension 
are divided into $L_r$ maximally overlapping subarrays, each containing $\subr{M} = M_r - L_r + 1$ 
elements. The corresponding $\subr{M} \times M_r$ selection matrix for the $\ell_r$-th subarray, 
$1 \leq \ell_r \leq L_r$ for $1 \leq r \leq R$, is defined as
\begin{align}
\bm J^{(M_r)}_{\ell_r} = \begin{bmatrix} \bm 0_{\subr{M} \times (\ell_r - 1)} & \bm I_{\subr{M}} 
& \bm 0_{\subr{M}\times(L_r - \ell_r)} \end{bmatrix}.
\label{selectsub}
\end{align}
Next, we define the $L = \prod_{r=1}^R L_r$ multi-dimensional selection matrices 
\begin{align} 
\bm J_{\underline{\bm \ell}} & = \bm J_{\ell_1,\ldots,\ell_{R-1},\ell_R} \notag\\
& = \bm J^{(M_1)}_{\ell_1}  \otimes  \cdots  \otimes 
\bm J^{(M_{R-1})}_{\ell_{R-1}}  \otimes  \bm J^{(M_R)}_{\ell_R} \!\! \in \! \real^{\sub{M} \times M} 
\label{selectrd}
\end{align}
%
%
for $1 \leq \ell_r \leq L_r$ with $\sub{M} = \prod_{r=1}^R \subr{M}$. Then, the spatially smoothed 
data matrix $\ssm{\bm X} \in\compl^{\sub{M}\times NL}$, which is subsequently processed instead of 
$\bm X$, is given by
\begin{align}
\ssm{\bm X} & = \left[\begin{matrix} \bm J_{1,\cdots,1,1} \bm X & \bm J_{1,\cdots,1,2} \bm X 
& \cdots & \bm J_{1,\cdots,1,L_R} \bm X \end{matrix} \right. \notag \\
& \left. \quad~\: \begin{matrix}  \bm J_{1,\cdots,2,1} \bm X & \cdots 
& \bm J_{L_1,\cdots,L_{R-1},L_R} \bm X \end{matrix} \right] \notag \\
& = \left[\begin{matrix} \bm J_{1,\cdots,1,1} \bar{\bm A} \bar{\bm S} & \cdots 
& \bm J_{L_1,\cdots,L_{R-1},L_R} \bar{\bm A} \bar{\bm S} \end{matrix} \right] \notag \\
& \quad~ + \left[\begin{matrix} \bm J_{1,\cdots,1,1} \bm N & \cdots 
& \bm J_{L_1,\cdots,L_{R-1},L_R} \bm N \end{matrix} \right]. 
\label{datass}  
\end{align}
Note that by using \eqref{steermat} and \eqref{selectrd}, the array steering matrix of the $\ell$-th 
subarray in all $R$ modes can be expressed as
\begin{align}
& \bm J_{\ell_1,\ldots,\ell_{R-1},\ell_R} \bar{\bm A} = \left( \bm J^{(M_1)}_{\ell_1} \bar{\bm A}^{(1)} \right) 
\diamond \cdots \diamond \left( \bm J^{(M_R)}_{\ell_R} \bar{\bm A}^{(R)} \right) \notag \\
& = \left( \bar{\bm A}_1^{(1)} (\bm \Phi^{(1)})^{\ell_1-1}\right) \diamond 
\cdots \diamond \left( \bar{\bm A}_1^{(R)}  (\bm \Phi^{(R)})^{\ell_R-1} \right) \notag \\
& = \ssm{\bar{\bm A}} \bm \Phi_{\ell_1,\ldots,\ell_{R-1},\ell_R},
\label{selonerd}
\end{align}
where we have defined
%
$\rd{\bar{\bm A}}_1 = \bm J^{(M_r)}_{1_r} \rd{\bar{\bm A}} \in\compl^{\subr{M} \times d}$, $\ssm{\bar{\bm A}} 
= \bar{\bm A}_1^{(1)} \diamond \cdots \diamond \bar{\bm A}_1^{(R)} = \bm J_{1,\cdots,1,1} \bar{\bm A} \in\compl^{\sub{M} \times d}$, 
and 
\begin{align}
\bm \Phi_{\ell_1,\ldots,\ell_{R-1},\ell_R} = \prod_{r = 1}^R (\bm \Phi^{(r)})^{\ell_r-1} . 
\notag
\end{align} 
%
Consequently, we can rewrite 
\eqref{datass} by applying \eqref{selonerd} as
\begin{align}
\ssm{\bm X} & = \ssm{\bar{\bm A}} \bm \Phi \left( \bm I_L \otimes \bar{\bm S} \right) + \ssm{\bm N} = \ssmnought{\bm X} + \ssm{\bm N} 
\label{smooth}
\end{align}
where $\bm \Phi = [\bm \Phi_{1,\ldots,1,1},\cdots,\bm \Phi_{1,\ldots,1,L_R},\bm \Phi_{1,\ldots,2,1},
\cdots,$ $\bm \Phi_{L_1,\ldots,L_{R-1},L_R}] \in\compl^{d \times Ld}$, $\ssmnought{\bm X} \in\compl^{
\sub{M}\times NL}$ is the noise-free spatially smoothed data matrix, and $\ssm{\bm N} \in\compl^{
\sub{M}\times NL}$ is the spatially smoothed noise. 
Thus, spatial smoothing preprocessing reduces the array aperture to $\sub{M}$ sensors and increases 
the number of snapshots 
by the factor $L$.

It is apparent that $\ssm{\bar{\bm A}}$ still satisfies the shift-invariance equation and we can write
\begin{align}
\ssmone{\rdtil{\bm J}} \ssm{\bar{\bm A}} ~\rd{\bm \Phi} = \ssmtwo{\rdtil{\bm J}} \ssm{\bar{\bm A}}, \quad r = 1,\ldots,R,
\label{shiftss}
\end{align}
where $\ssmone{\rdtil{\bm J}}$ and $\ssmtwo{\rdtil{\bm J}} \in\real^{\frac{\sub{M}}{\subr{M}} 
(\subr{M}-1) \times \sub{M}}$ are the $R$-D selection matrices that select $\subr{M}-1$ 
elements for the first and the second subarray in the $r$-th mode, respectively. 
They are compactly defined as $\ssmk{\rdtil{\bm J}} = \bm I_{\prod_{l=1}^{r-1} \subl{M}} \otimes \rd{\bm J}_{{\rm SS}_k} \otimes 
\bm I_{\prod_{l=r+1}^{R} \subl{M}}$ for $k=1,2$, where $\rd{\bm J}_{{\rm SS}_k} \in\real^{(\subr{M}-1) \times \subr{M}}$ are 
the $r$-mode selection matrices for the first and second subarray.
%
%
\red{
As \eqref{shiftss} holds, the $R {\cdot} d$ spatial frequencies can be estimated by applying $R$-D ESPRIT-type 
algorithms to $\ssm{\bm X}$. In $R$-D Standard ESPRIT, the signal subspace $\ssms{\hat{\bm U}} \in\compl^{\sub{M} 
\times d}$ is estimated by computing the $d$ dominant left singular vectors of $\ssm{\bm X}$. As 
$\ssm{\bar{\bm A}}$ and $\ssms{\hat{\bm U}}$ span approximately the same column space, a non-singular matrix 
$\bm T \in\compl^{d\times d}$ can be found such that $\ssm{\bar{\bm A}} \approx \ssms{\hat{\bm U}} \bm T$. 
Using this relation, the overdetermined set of $R$ shift invariance equations \eqref{shiftss} can be 
expressed in terms of the estimated signal subspace, yielding
\begin{align}
{\ssmone{\rdtil{\bm J}}} \ssms{\hat{\bm U}} \rd{\bm \Gamma} \approx {\ssmtwo{\rdtil{\bm J}}} \ssms{\hat{\bm U}}, \quad r=1,\ldots,R
\label{shiftsubs}
\end{align}
with $\rd{\bm \Gamma} = \bm T \rd{\bm \Phi} \bm T^{-1}$. The $R$ unknown matrices $\rd{\bm \Gamma} \in\compl^{d\times d}$ 
can be estimated, e.g., via least squares (LS), i.e.,
\begin{equation}
\rd{\hat{\bm \Gamma}} = \left({\ssmone{\rdtil{\bm J}}} \ssms{\hat{\bm U}} \right)^+ {\ssmtwo{\rdtil{\bm J}}} \ssms{\hat{\bm U}}  
\in\compl^{d\times d}.
\label{lsstan}
\end{equation}
Finally, after solving \eqref{lsstan} for $\rd{\hat{\bm \Gamma}}$ in each mode independently, the correctly paired 
spatial frequency estimates are given by $\rd{\hat{\mu}}_i=\mathrm{arg}\{\rd{\hat{\lambda}}_i\},~i=1,\ldots,d$. 
The eigenvalues $\rd{\hat{\lambda}}_i$ of $\rd{\hat{\bm \Gamma}}$ are obtained by performing a joint 
eigendecomposition across all $R$ dimensions \cite{fu2006diag} or via the simultaneous Schur decomposition 
\cite{haardt1998schur}.
Alternatively, $R$-D Unitary ESPRIT \cite{haardt1998schur} can be applied to estimate the $R {\cdot} d$ parameters, which 
is preferable due to its better performance at low SNRs and its real-valued implementation. 
}
\subsection{$R$-D Spatial Smoothing for Strictly Non-Circular Sources}
\label{sec:noncirc}
If only NC sources are present, a modified spatial smoothing concept can be applied to the NC model in \eqref{ncmodel} \cite{haardt2004ncunit}, where 
we select $2 \sub{M}$ out of $2M$ virtual sensors. Thus, the $L$ selection matrices in \eqref{selectrd} 
are extended to 
\begin{align}
\nc{\bm J}_{\ell_1,\ldots,\ell_{R-1},\ell_R} = \bm I_2 \otimes \bm J_{\ell_1,\ldots,\ell_{R-1},\ell_R} \in\real^{2\sub{M} \times 2M}.
\label{smoothncsel}
\end{align}
The resulting spatially smoothed data matrix $\ssmnc{\bm X}$ of size $2\sub{M}\times NL$ is then given 
by
\begin{align}
\ssmnc{\bm X} & = \left[\begin{matrix} \nc{\bm J}_{1,\cdots,1,1} \nc{\bm X}  
& \cdots & \nc{\bm J}_{1,\cdots,1,L_r} \nc{\bm X} \end{matrix} \right. \notag \\
& \left.  \begin{matrix}  \nc{\bm J}_{1,\cdots,2,1} \nc{\bm X} & \cdots 
& \nc{\bm J}_{L_1,\cdots,L_{R-1},L_R} \nc{\bm X} \end{matrix} \right].  
\label{smoothnc}
\end{align}
Following the lines of the previous subsection, we can compactly express \eqref{smoothnc} as 
\begin{align}
\ssmnc{\bm X} & = \ssmnc{\bar{\bm A}} \bm \Phi \left( \bm I_L \otimes \bar{\bm S} \right) + \ssmnc{\bm N} \notag \\
& = \nc{\ssmnought{\bm X}} + \ssmnc{\bm N} ~\in\compl^{2\sub{M}\times NL},
\label{smoothnc2}
\end{align}
where 
%
$\ssmnc{\bar{\bm A}} = \nc{\bm J}_{1,\cdots,1,1} \nc{\bar{\bm A}} \in\compl^{2\sub{M} \times d}$
%
and $\nc{\ssmnought{\bm X}}$ is the unperturbed spatially smoothed NC data matrix. Note that spatial 
smoothing cannot be applied before $\nc{\bm X}$ is formed \eqref{ncmodel} as this would destroy 
the NC structure of the source signals.

As in the previous cases, $\ssmnc{\bm A}$ is shift-invariant and satisfies
\begin{equation}
\ssmone{\ncrtil{\bm J}} \ssmnc{\bar{\bm A}} \rd{\bm \Phi} = \ssmtwo{\ncrtil{\bm J}} \ssmnc{\bar{\bm A}}, \quad r=1,\ldots,R,
\label{shiftssnc}
\end{equation}
where $\ssmk{\ncrtil{\bm J}} \in\real^{2\frac{\sub{M}}{\subr{M}} \subsel{\subr{M}} \times 2\sub{M}},~k=1,2$ 
are the corresponding selection matrices that select $2\subsel{\subr{M}}$ elements for the first and the 
second subarray in the $r$-th mode. They are defined as $\ssmk{\ncrtil{\bm J}} = \bm I_{\prod_{l=1}^{r-1} M_l} 
\otimes \ncr{\bm J}_{{\rm SS}_k} \otimes \bm I_{\prod_{l=r+1}^{R} M_l}$, where $\ncr{\bm J}_{{\rm SS}_k} = 
\bm I^{}_2 \otimes \rd{\bm J}_{{\rm SS}_k} \in\real^{2\subsel{\subr{M}} \times 2\subr{M}}$ are 
the $r$-mode selection matrices for the first and second subarray.
Again, $R$-D ESPRIT-type algorithms such as $R$-D NC Standard ESPRIT and $R$-D NC Unitary ESPRIT \cite{steinwandt2014tsp} 
can be used to estimate the $R {\cdot} d$ parameters.
\section{Performance of $R$-D ESPRIT-Type Algorithms with Spatial Smoothing}
In this section, we present first-order error expansions of $R$-D Standard ESPRIT and $R$-D Unitary 
ESPRIT both with spatial smoothing. The derived expressions rely on the data model \eqref{smooth} in 
Section \ref{sec:circ}.
\subsection{R-D Standard ESPRIT with Spatial Smoothing}
\label{sec:SESS}
For the perturbation analysis of the estimation error, we adopt the analytical framework proposed 
in \cite{vaccaro1993perf} along with its extension in \cite{roemer2014perfana}. The authors of 
\cite{vaccaro1993perf} assume a small additive noise perturbation and derive an explicit first-order 
error expansion of the subspace estimation error in terms of the noise $\bm N$, which is followed by 
a corresponding expression for the parameter estimation error $\Delta \mu_i$. As a follow-up, 
analytical expressions for the MSE that only require a zero mean and finite SO moments of the noise 
have been derived in \cite{roemer2014perfana}. From \eqref{smooth}, it is clear that these assumptions 
are not violated by 
spatial smoothing such that \cite{vaccaro1993perf} and 
\cite{roemer2014perfana} are still applicable for the performance analysis. 

To derive the signal subspace estimation error for \eqref{smooth}, we express the SVD of the 
noise-free spatially smoothed observations $\ssmnought{\bm X}$ as
\begin{align}
\ssmnought{\bm X} = \begin{bmatrix} \ssms{\bm U} & \ssmn{\bm U} \end{bmatrix}
\begin{bmatrix} \ssms{\bm \Sigma} & \bm 0 \\ \bm 0 & \bm 0 \end{bmatrix}
\begin{bmatrix} \ssms{\bm V} & \ssmn{\bm V} \end{bmatrix}^\herm,
\label{noisefree}
\end{align}
where $\ssms{\bm U} \in\compl^{\sub{M}\times d}$, $\ssmn{\bm U} \in\compl^{\sub{M} \times (NL-d)}$, 
and $\ssms{\bm V} \in\compl^{NL \times d}$ span the signal subspace, the noise subspace, and 
the row space, respectively, and $\ssms{\bm \Sigma} \in\real^{d\times d}$ contains the non-zero 
singular values on its diagonal. Writing the perturbed signal subspace estimate $\ssms{\hat{\bm U}}$ 
computed from the SVD of $\ssm{\bm X}$ as $\ssms{\hat{\bm U}} = \ssms{\bm U} + \Delta \ssms{\bm U}$, 
where $\Delta \ssms{\bm U}$ denotes the signal subspace error, the first-order approximation 
using \cite{vaccaro1993perf} is given by
\begin{align}
\Delta \ssms{\bm U} = \ssmn{\bm U}^{} \ssmn{\bm U}^\herm \ssm{\bm N}^{} \ssms{\bm V}^{} \ssms{\bm \Sigma}^{-1} + \red{\mathcal O\{\nu^2\}},
\label{subpert}
\end{align}
where \red{$\nu = \|\ssm{\bm N}\|$,} and $\|\cdot\|$ represents an arbitrary sub-multiplicative norm\footnote{A 
matrix norm is called sub-multiplicative if $\|\bm A\cdot \bm B \|\leq \|\bm A\|\cdot \|\bm B\|$ for 
arbitrary matrices $\bm A$ and $\bm B$.}. \red{For the estimation error $\Delta \rd{\mu}_i$ of the $i$-th spatial frequency 
in the $r$-th mode obtained by the LS solution, we have} \cite{vaccaro1993perf}
%
%
\begin{eqnarray}
\begin{aligned}
\Delta \rd{\mu}_i & = \mathrm{Im}\left\{\bm p_i^\trans \left( \ssmone{\rdtil{\bm J}} \ssms{\bm U} \right)^+
\left[\ssmtwo{\rdtil{\bm J}}/ \rd{\lambda}_i \right.\right.\\
&\qquad~~\qquad \left.\left. - \ssmone{\rdtil{\bm J}} \right] \Delta \ssms{\bm U} \bm q_i \right\} + \red{\mathcal O\{\nu^2\}},
\label{estpert}
\end{aligned}
\end{eqnarray}
where $\rd{\lambda}_i = \e^{\j \rd{\mu}_i}$ is the $i$-th eigenvalue of $\rd{\bm \Gamma}$, $\bm q_i$ represents 
the $i$-th eigenvector of $\rd{\bm \Gamma}$ and the $i$-th column vector of the eigenvector matrix $\bm Q$, 
and $\bm p_i^\trans$ is the $i$-th row vector of $\bm P = \bm Q^\inv$. Hence, the eigendecomposition of 
$\rd{\bm \Gamma}$ is given by $\rd{\bm \Gamma} = \bm Q \rd{\bm \Lambda} \bm Q^\inv$, where $\rd{\bm \Lambda}$ 
contains the eigenvalues $\rd{\lambda}_i$ on its diagonal. 

Finally, to compute the first-order MSE expression for $R$-D Standard ESPRIT with spatial smoothing, we extend 
the results in \cite{roemer2014perfana}. 
The MSE for the $i$-th spatial frequency in the $r$-th mode is given by
\begin{eqnarray}
\begin{aligned}
\mathbb E\left\{(\Delta \rd{\mu}_i)^2\right\} & \approx \frac{1}{2} \left( \rdH{\ssmi{\bm r}} \ssm{\bm W}^\conj
\ssm{\bm R}^\trans \ssm{\bm W}^\trans {\rd{\ssmi{\bm r}}} \right.\\
&\left. \quad~~ -\mathrm{Re}\left\{ \rdT{\ssmi{\bm r}} \ssm{\bm W}
\ssm{\bm C}^\trans \ssm{\bm W}^\trans {\rd{\ssmi{\bm r}}} \right\}\right),
\label{mse}
\end{aligned}
\vspace{-1em}
\end{eqnarray}
where
\begin{align}\notag
{\rd{\ssmi{\bm r}}}  = \bm q_i \otimes \left(\left[\left(\ssmone{\rdtil{\bm J}} \ssms{\bm U} \right)^+
\left(\ssmtwo{\rdtil{\bm J}} / \rd{\lambda}_i - \ssmone{\rdtil{\bm J}} \right)\right]^\trans \bm p_i\right),
\end{align}
\vspace{-1em}
\begin{align}\notag
\ssm{\bm W} = \left( \ssms{\bm \Sigma}^{-1} \ssms{\bm V}^\trans \right) 
\otimes \left( \ssmn{\bm U} \ssmn{\bm U}^\herm \right) \in\compl^{\sub{M}d \times \sub{M} N L}.
\end{align}
In order to apply \eqref{mse}, we require the covariance matrix $\ssm{\bm R} = \mathbb E\{ \ssm{\bm n}^{} \ssm{\bm n}^\herm\} 
\in\compl^{\sub{M}NL \times \sub{M}NL}$ and the pseudo-covariance matrix $\ssm{\bm C} = \mathbb E\{ 
\ssm{\bm n}^{} \ssm{\bm n}^\trans\} \in\compl^{\sub{M}NL \times \sub{M}NL}$ of the spatially 
smoothed noise $\ssm{\bm n} = \mathrm{vec}\{\ssm{\bm N}\} \in\compl^{\sub{M}NL \times 1}$. It is clear 
that the preprocessing via spatial smoothing 
modifies the prior noise statistics, resulting in colored noise. 
However, in what follows, we analytically derive the SO noise statistics of the spatially smoothed noise.
We first expand $\ssm{\bm n}^{}$ as
\begin{align}
\ssm{\bm n}^{} & = \vecof{\begin{bmatrix} \bm J_{1,\cdots,1,1} \bm N & \cdots & \bm J_{L_1,\cdots,L_{R-1},L_R} \bm N \end{bmatrix}} \notag \\
& = \begin{bmatrix} (\bm I_N \otimes \bm J_{1,\cdots,1,1}) \\ \vdots \\ (\bm I_N \otimes \bm J_{L_1,\cdots,L_{R-1},L_R}) \end{bmatrix} 
\cdot \bm n = \red{\bm M \cdot \bm n},
\label{nss2}
\end{align}
where $\red{\bm M} 
\in\real^{\sub{M}NL \times MN}$, $\bm n = \vecof{ \bm N } \in\compl^{MN\times 1}$ is the unsmoothed noise component, 
and we have used the property $\vecof{\bm A \bm X \bm B } = (\bm B^\trans \otimes \bm A) \vecof{\bm X }$ for arbitrary 
matrices $\bm A$, $\bm B$, and $\bm X$ of appropriate sizes. 
Thus, the SO statistics of $\ssm{\bm n}$ can be expressed in terms of the covariance matrix $\bm R_\mathrm{nn} 
= \mathbb E\{\bm n \bm n^\herm\} \in\compl^{MN \times MN}$ and the pseudo-covariance matrix $\bm C_\mathrm{nn} 
= \mathbb E\{\bm n \bm n^\trans\} \in\compl^{MN \times MN}$ of $\bm n$. We obtain
%
\red{
\begin{align}
\ssm{\bm R} & = \bm M \bm R_\mathrm{nn} \bm M^\trans, \quad
\ssm{\bm C} = \bm M \bm C_\mathrm{nn} \bm M^\trans. 
\label{covss}
\end{align} }
%
%
%
\subsection{$R$-D Unitary ESPRIT with Spatial Smoothing}
It was shown in \cite{roemer2014perfana} that the asymptotic performance of $R$-D Unitary-ESPRIT is 
found once forward-backward-averaging (FBA) is taken into account. FBA is performed by replacing the 
spatially smoothed data matrix $\ssm{\bm X} \in\compl^{\sub{M} \times NL}$ by the column-augmented 
data matrix $\ssmtil{\bm X} \in\compl^{\sub{M} \times 2NL}$ defined by
\begin{align}
\ssmtil{\bm X} & = \begin{bmatrix} \ssm{\bm X} & \bm \Pi_{\sub{M}} \ssm{\bm X}^\conj \bm \Pi_{NL} \end{bmatrix} 
= \ssmnought{\tilde{\bm X}} + \ssmtil{\bm N}, \label{fba}  
\end{align}
where $\ssmnought{\tilde{\bm X}}$ is the noiseless FBA-processed spatially smoothed data matrix. 
Following the steps of the previous subsection, the first-order MSE expression for $R$-D Unitary ESPRIT 
with spatial smoothing for the $i$-th spatial frequency in the $r$-th mode is given by
\begin{eqnarray}
\begin{aligned}
\mathbb E\left\{(\Delta \rd{\mu}_i)^2\right\} & \approx \frac{1}{2} \left( \ssmi{\rdtilH{\bm r}} \ssm{\tilde{\bm W}}^\conj
\ssm{\tilde{\bm R}}^\trans \ssm{\tilde{\bm W}}^\trans {\rd{\ssmi{\tilde{\bm r}}}} \right.\\
&\left. \quad~~ -\mathrm{Re}\left\{ \ssmi{\rdtilT{\bm r}} \ssm{\tilde{\bm W}}
\ssm{\tilde{\bm C}}^\trans \ssm{\tilde{\bm W}}^\trans {\rd{\ssmi{\tilde{\bm r}}}} \right\}\right)
\label{mse_fba}
\end{aligned}
\vspace{-1em}
\end{eqnarray}
with
\begin{align}\notag
{\rd{\ssmi{\tilde{\bm r}}}}  = \tilde{\bm q}_i \otimes \left(\left[\left(\ssmone{\rdtil{\bm J}} \ssmtils{\bm U} \right)^+
\left(\ssmtwo{\rdtil{\bm J}} / \rd{\lambda}_i - \ssmone{\rdtil{\bm J}} \right)\right]^\trans \tilde{\bm p}_i \right),
\end{align}
\vspace{-1em}
\begin{align}\notag
\ssm{\tilde{\bm W}} = \left( \ssmtilsinv{\bm \Sigma} {\ssmtils{\bm V}}^\trans\right) 
\otimes \left( \ssmtiln{\bm U}^{} \ssmtiln{\bm U}^\herm \right) \in\compl^{\sub{M}d \times 2\sub{M}NL},
\end{align}
where we have replaced the noise-free subspaces of $\ssmnought{\bm X}$ in \eqref{mse} by the corresponding 
subspaces of $\ssmnought{\tilde{\bm X}}$, and $\bm p_i$ and $\bm q_i$ by $\tilde{\bm p}_i$ and $\tilde{\bm q}_i$, 
respectively. 
%
%
It can be shown 
that $\ssmtil{\bm n} = \mathrm{vec}\{ \ssmtil{\bm N} \} \in\compl^{2\sub{M}NL \times 1}$ is given by
\begin{align}
\ssmtil{\bm n} & = \vecof{\begin{bmatrix} \ssm{\bm N} & \bm \Pi_{\sub{M}} \ssm{\bm N}^\conj \bm \Pi_{NL} \end{bmatrix} } \notag \\
& = \begin{bmatrix} \vecof{\ssm{\bm N}} \\ \vecof{\bm \Pi_{\sub{M}} \ssm{\bm N}^\conj \bm \Pi_{NL}} \end{bmatrix}  
 = \begin{bmatrix} \ssm{\bm n} \\ \bm \Pi^{}_{{\sub{M}}NL} \ssm{\bm n}^\conj \end{bmatrix}.
\label{noisessfba}
\end{align}
Therefore, the expressions for $\ssmtil{\bm R} = \mathbb E\{\ssmtil{\bm n}^{} \ssmtil{\bm n}^\herm\} 
\in\compl^{2\sub{M}NL \times 2\sub{M}NL}$ and $\ssmtil{\bm C} = \mathbb E\{\ssmtil{\bm n}^{} 
\ssmtil{\bm n}^\trans\} \in\compl^{2\sub{M}NL \times 2\sub{M}NL}$ can be derived in terms of 
\eqref{covss} as
\begin{align} \notag
\ssmtil{\bm R} = \bm P\begin{bmatrix} \ssm{\bm R} & \ssm{\bm C} \\
\ssm{\bm C}^\conj & \ssm{\bm R}^\conj \end{bmatrix} \bm P^\trans, ~~
\ssmtil{\bm C} = \bm P\begin{bmatrix} \ssm{\bm C} & \ssm{\bm R} \\ 
\ssm{\bm R}^\conj & \ssm{\bm C}^\conj \end{bmatrix} \bm P^\trans,
\end{align}
where $\bm P = \mathrm{blkdiag}\{\bm I^{}_{\sub{M}NL},~\bm \Pi^{}_{{\sub{M}}NL}\}$.
\section{Performance of $R$-D NC ESPRIT-Type Algorithms with Spatial Smoothing}
\label{sec:perf}
In this section, we derive first-order analytical error expressions of $R$-D NC Standard ESPRIT and $R$-D NC Unitary 
ESPRIT for strictly non-circular sources both with spatial smoothing. As will be shown in Subsection \ref{sec:ncunit}, 
the performance of both algorithms is asymptotically identical in the high effective SNR. Therefore, we first resort 
to the simpler derivation for the spatially smoothed $R$-D NC Standard ESPRIT algorithm and then show its equivalence to the spatially smoothed $R$-D NC Unitary ESPRIT algorithm. 
Our results are based on the data model \eqref{smoothnc2} in Section \ref{sec:noncirc}. 
\subsection{$R$-D NC Standard ESPRIT with Spatial Smoothing}
%
%
In \cite{steinwandt2014tsp}, we have shown that the framework of \cite{vaccaro1993perf} is still applicable 
to the augmented measurement matrix $\bm X^{(\mathrm{nc})}$ \eqref{ncmodel} obtained by the preprocessing 
scheme for non-circular sources. From \eqref{smoothnc2}, it is apparent that adding spatial smoothing as a 
second preprocessing step does not violate the assumptions, such that the steps from Section \ref{sec:SESS} 
can be applied to the spatially smoothed augmented data matrix $\nc{\bm X}_{\rm SS}$. 

As a result, equivalently to \eqref{mse}, the first-order MSE expression for $R$-D NC Standard ESPRIT 
with spatial smoothing for the $i$-th spatial frequency in the $r$-th mode is given by
%
\begin{align}
&\mathbb E\left\{(\Delta \rd{\mu}_i)^2\right\} \approx \frac{1}{2} \left( \ncrH{\ssmi{\bm r}} \ncC{\ssm{\bm W}}
\ncT{\ssm{\bm R}} \ncT{\ssm{\bm W}} \ncr{\ssmi{\bm r}} \right. \notag\\
&\left. \quad\quad -\mathrm{Re}\left\{ \ncrT{\ssmi{\bm r}} \nc{\ssm{\bm W}}
\ncT{\ssm{\bm C}} \ncT{\ssm{\bm W}} \ncr{\ssmi{\bm r}} \right\}\right), 
\label{mse_nc}
\end{align}
%
where
\begin{align}\notag
\ssmi{\ncrtil{\bm r}} & = \nc{\bm q}_i \otimes \left(\left[\left(\ssmone{\ncrtil{\bm J}} \ssmncs{\bm U} \right)^+ \right. \right. \\
& \left. \left. \cdot \left(\ssmtwo{\ncrtil{\bm J}} / \rd{\lambda}_i - \ssmone{\ncrtil{\bm J}} \right)\right]^\trans \nc{\bm p}_i \right) 
\in\compl^{2\sub{M}d \times 1},
\notag
\end{align}
%
%
%
\small
\begin{align} \notag
\nc{\ssm{\bm W}} \!\! = \! \left( \ssmncsinv{\bm \Sigma} \!\! \ssmncsT{\bm V} \right) \! \otimes \! 
\left( \ssmncn{\bm U} \ssmncnH{\bm U} \right) \! \in\compl^{2\sub{M}d \times 2\sub{M}NL},
\end{align}
\normalsize
where $\nc{\bm p}_i$ and $\nc{\bm q}_i$ replace $\bm p_i$ and $\bm q_i$, respectively, we have used the 
corresponding subspaces of $\nc{\ssmnought{\bm X}}$ defined in \eqref{smoothnc2}, and the selection 
matrices $\ssmk{\ncrtil{\bm J}}, k=1,2$, are given in \eqref{shiftssnc}.


The spatially smoothed augmented noise contribution $\ssmnc{\bm n} = \mathrm{vec}\{\ssmnc{\bm N}\} 
\in\compl^{2\sub{M}NL\times 1}$ can be expressed similarly to \eqref{nss2} as
\begin{align}
\ssmnc{\bm n} & \! = \vecof{\begin{bmatrix} \nc{\bm J}_{1,\cdots,1,1} \nc{\bm N} & \! \cdots  \! & 
\nc{\bm J}_{L_1,\cdots,L_{R-1},L_R} \nc{\bm N} \end{bmatrix}} \notag \\
& \!\!\!\!\! = \! \begin{bmatrix} (\bm I_N \otimes \nc{\bm J}_{1,\cdots,1,1}) \\ \vdots \\ (\bm I_N \otimes \nc{\bm J}_{L_1,\cdots,L_{R-1},L_R}) \end{bmatrix} \!
\cdot \nc{\bm n} \! = \red{\nc{\bm M} \! \cdot \nc{\bm n} \! },
\label{noisessnc}
\end{align}
where $\red{\nc{\bm M}}
\in\real^{2\sub{M}NL \times 2MN}$ and $\nc{\bm n} = \mathrm{vec}\{\nc{\bm N}\} \in\compl^{2MN\times 1}$. 
Note that we have shown in \cite{steinwandt2014tsp} that $\nc{\bm n}$ can be represented as
\begin{align}
\nc{\bm n} = \tilde{\bm K} \cdot \begin{bmatrix} \bm n \\ \bm n^\conj \end{bmatrix},
\end{align}
where $\tilde{\bm K} = \bm K_{2M,N}^\trans \cdot \mathrm{blkdiag}\{\bm K_{M,N}$,\,$\bm K_{M,N} \cdot
\left(\bm I_{N} \otimes \bm \Pi_M \right)\}$ and $\bm K_{M,N} \in \real^{MN\times MN}$ is the commutation 
matrix that satisfies $\bm K_{M,N}\cdot\mathrm{vec}\{\bm A\} = \mathrm{vec} \{\bm A^\trans\}$ for arbitrary 
matrices $\bm A \in\mathbb C^{M\times N}$ \cite{magnus1995matrix}. Then, $\ssmnc{\bm R} = \mathbb E\{\ssmnc{\bm n} 
\ssmncH{\bm n}\} \in\compl^{2\sub{M}NL \times 2\sub{M}NL}$ and $\ssmnc{\bm C} = \mathbb E\{\ssmnc{\bm n} 
\ssmncT{\bm n}\} \in\compl^{2\sub{M}NL \times 2\sub{M}NL}$ can be computed as
%
\red{
\begin{align} 
\ssmnc{\bm R} & = \nc{\bm M} \nc{\bm R}_\mathrm{nn} \ncT{\bm M}, \quad
\ssmnc{\bm C} = \nc{\bm M} \nc{\bm C}_\mathrm{nn} \ncT{\bm M}, 
\label{covssnc}
\end{align}
}
where $\nc{\bm R}_\mathrm{nn} \in\compl^{2MN \times 2MN}$ and $\nc{\bm C}_\mathrm{nn} \in\compl^{2MN \times 2MN}$ are given 
by \cite{steinwandt2014tsp}
\begin{align} 
\nc{\bm R}_\mathrm{nn} & = \mathbb E \left\{ \nc{\bm n} \ncH{\bm n} \right\} = \tilde{\bm K} \begin{bmatrix} \bm R_{\rm nn} & \bm C_{\rm nn} \\
\bm C_{\rm nn}^\conj & \bm R_{\rm nn}^\conj \end{bmatrix} \tilde{\bm K}^\trans, \\
\nc{\bm C}_\mathrm{nn} & = \mathbb E \left\{ \nc{\bm n} \ncT{\bm n} \right\} = \tilde{\bm K} \begin{bmatrix} \bm C_{\rm nn} & \bm R_{\rm nn} \\ 
\bm R_{\rm nn}^\conj & \bm C_{\rm nn}^\conj \end{bmatrix} \tilde{\bm K}^\trans.
\end{align}
\subsection{$R$-D NC Unitary ESPRIT with Spatial Smoothing}
\label{sec:ncunit}
%
%
We have shown in \cite{steinwandt2014tsp} that $R$-D NC Standard ESPRIT and $R$-D NC Unitary ESPRIT 
both have the same asymptotic performance in the high effective SNR regime. It was established 
that applying FBA to the augmented matrix $\nc{\bm X}$ does not improve the signal subspace 
estimate and that the real-valued transformation has no effect on the asymptotic performance in the high 
effective SNR. In this subsection, we prove that these properties still hold 
when spatial smoothing 
is applied to both algorithms. 
To this end, we first investigate the effect of FBA and state the following theorem:
\begin{thm}
Applying FBA to $\ssmnc{\bm X}$ does not improve the signal subspace estimate.
\label{thm:fba}
\end{thm}
\begin{IEEEproof}
The proof is given in Appendix \ref{app:fba}.
\end{IEEEproof}
Next, we analyze the real-valued transformation as the second preprocessing step of $R$-D NC Unitary 
ESPRIT with spatial smoothing and formulate the theorem:
\begin{thm}
The spatially smoothed $R$-D NC Unitary ESPRIT algorithm and the spatially smoothed $R$-D NC Standard 
ESPRIT algorithm with FBA preprocessing perform asymptotically identical in the high effective SNR.
\label{thm:realtrans}
\end{thm}
\begin{IEEEproof} 
The proof of this theorem follows the same steps as the one for the case without spatial smoothing considered 
in \cite{steinwandt2014tsp}. This is due to the fact that spatial smoothing modifies the NC signal subspace 
of $R$-D NC Standard ESPRIT and $R$-D NC Unitary ESPRIT in the same way.
\end{IEEEproof}
As a result of Theorem \ref{thm:fba} and Theorem \ref{thm:realtrans}, we can conclude that the asymptotic 
performance of $R$-D NC Standard ESPRIT and $R$-D NC Unitary ESPRIT both with spatial smoothing is identical 
in the high effective SNR. 
\section{Single Source Case}
\label{sec:single}
The derived analytical MSE expressions for the $R$-D ESPRIT-type methods with spatial smoothing are deterministic and 
formulated in terms of the subspaces of the noise-free observations. In \cite{roemer2012perf} and \cite{steinwandt2014tsp}, 
we have considered the special case of a single source for $R$-D ESPRIT-type methods without spatial smoothing to gain 
explicit insights into how the MSE expressions depend on the physical parameters, e.g., the number of sensors $M$, the 
sample size $N$, and the SNR. The knowledge of how the MSE expressions depend on these parameters can be of practical 
significance. For instance, this enables an objective comparison of different estimators or facilitates array design 
decisions on the value of $M$ required to achieve a target MSE for a specific SNR. Note that establishing general MSE 
expressions for an arbitrary number of sources is challenging given the complex dependence of the subspaces on the 
physical parameters. For the single source case, it was proven in \cite{roemer2012perf} and \cite{steinwandt2014tsp} that 
neither FBA nor NC preprocessing can improve the MSE. However, in this section, we show that a significant gain can be 
achieved for the MSE of $R$-D ESPRIT-type methods for a single source when spatial smoothing is applied. Assuming 
an $R$-D uniform sampling grid, i.e., a ULA in each mode, and circularly symmetric white noise, we simplify the derived 
MSE expressions in \eqref{mse}, \eqref{mse_fba}, and \eqref{mse_nc} for this special case. The result depends on the 
number of subarrays $L_r$ in the $r$-th mode as a design parameter, which we analytically compute in the $R$-D case by
minimizing the MSE. It should be emphasized that these results for the special case $R=1$ are in line with those derived 
in \cite{rao1990spasmoo, rao1993spasmoo,hua1990spasmoo} for harmonic retrieval. Here, the $R$-D extension is provided. 
Based on our $R$-D results, we explicitly compute the asymptotic spatial smoothing gain for arbitrary $R$ and the asymptotic 
efficiency for $R=1$ in closed-form. 
%
%
%
%
\subsection{R-D ESPRIT-type Algorithms with Spatial Smoothing}
The final result for the simplified MSE expressions is summarized in the following theorem:
\begin{thm}
For the case of an $M$-element $R$-D uniform sampling grid with an $M_r$-element ULA in the $r$-th mode, a 
single source ($d=1$), and circularly symmetric white noise, the MSE in the $r$-th mode of $R$-D Standard ESPRIT 
and $R$-D Unitary ESPRIT with spatial smoothing as well as the MSE in the $r$-th mode of $R$-D NC Standard ESPRIT 
and $R$-D NC Unitary ESPRIT with spatial smoothing for a single 
source are given by 
$\rd{{\rm MSE}}_{\rm SS} = \mathbb E\left\{(\Delta\rd{\mu})^2\right\}$, yielding
%
%
\small
\begin{align} 
\rd{{\rm MSE}}_{\rm SS} \approx \!
\begin{cases} 
		\frac{1}{\hat{\rho}} \cdot \! \frac{1}{(M_r - L_r)^2 L_r} \cdot \! \prod_{\substack{p = 1 \\ p \neq r}}^R \frac{c_p}{\subp{M}^2 L_p^2}
& \text{if } L_r \leq \frac{M_r}{2} \\
    \frac{1}{\hat{\rho}} \cdot \! \frac{1}{(M_r - L_r) L_r^2} \cdot \! \prod_{\substack{p = 1 \\ p \neq r}}^R \frac{c_p}{\subp{M}^2 L_p^2}
    & \text{if } L_r > \frac{M_r}{2},
\end{cases}
\label{msespasmoosingle}
\end{align}
\normalsize
where $c_p$ is given in \eqref{cp} 
\begin{figure*}[bp]
\hrule
\small
\begin{align}
c_p = \frac{1}{3} \cdot \Big( \min\{ L_p, M_p - L_p \} + 1 \Big) \Big( \min\{ L_p, M_p - L_p \} 
\big( 2 \cdot \min\{ L_p, M_p - L_p \} - 3 \cdot M_p - 2 \big) + 6 \cdot \subp{M} L_p \Big) - \subp{M} L_p 
\label{cp}
\end{align}
\end{figure*}
\normalsize
and $\hat{\rho}$ represents the effective SNR $\hat{\rho} = N \hat{P}_\mathrm{s} 
/\sigman^2$ with $\hat{P}_\mathrm{s}$ being the empirical source power given by $\hat{P}_\mathrm{s} = \twonorm{\bm s}^2/N$ 
and $\bm s \in \compl^{N \times 1}$. 
\label{thm:ncunit_single}
\end{thm}
\begin{IEEEproof}
See Appendix \ref{app:ncunit_single}.
\end{IEEEproof}
Note that \eqref{msespasmoosingle} as a function of $L_r$ is symmetric with respect to $L_r = M_r/2$.  
In the special case of $R=1$, where $M_r = M$ and $L_r = L$, the MSE in \eqref{msespasmoosingle} 
simplifies to
\begin{align} 
{\rm MSE_{SS}} \approx 
\begin{cases} \frac{1}{\hat{\rho}} \cdot \frac{1}{(M-L)^2L} 
& \text{if } L \leq \frac{M}{2} \\
\frac{1}{\hat{\rho}} \cdot \frac{1}{(M-L)L^2} 
& \text{if } L > \frac{M}{2}.
\end{cases}
\label{msespasmoosingle_1d}
\end{align}
Interestingly, we arrive at the same result for the MSE of all the considered spatially smoothed $R$-D ESPRIT-type 
algorithms for a single source, i.e., no additional gain from FBA or NC preprocessing can be achieved. 
\subsection{Optimal Number of Subarrays for Spatial Smoothing}
\label{sec:optl}
In the MSE expression in \eqref{msespasmoosingle}, the number of subarrays $L_r$ in each mode is a design parameter that 
can be optimized. Therefore, minimizing the MSE expression \eqref{msespasmoosingle} with respect to $L_r$, yields\footnote{As 
\eqref{msespasmoosingle} is symmetric with respect to $L_r = M_r/2$, we obtain two values for $L_r^{\rm opt}$ that both 
minimize the MSE and are equally valid. }
\begin{align}
L_r^{\rm opt} = \begin{cases} \frac{1}{3} \cdot M_r & \text{if } L_r \leq \frac{M_r}{2} \\ 
\frac{2}{3} \cdot M_r & \text{if } L_r > \frac{M_r}{2},  \end{cases}
\label{lopt}
\end{align}
where it is assumed that $M_r$ is a multiple of $3$. \red{A short proof is provided in Appendix \ref{app:lopt}. }
%
%
If $M_r$ is not a multiple of $3$, we round to the nearest integer. Then, $L_r^{\rm opt}$ for the case $L_r 
\leq \frac{M_r}{2}$, for instance, is given by 
\begin{align}
L_r^{\rm opt} = \begin{cases} \frac{1}{3} \cdot (M_r - 1) & \text{if } M_r~ {\rm mod}~3 = 1 \\ 
\frac{1}{3} \cdot (M_r + 1) & \text{if } M_r~{\rm mod}~3 = 2. \end{cases}
\label{loptmod}
\end{align}
\red{It is worth highlighting that $L_r^{\rm opt}$ is independent of $L_p$ and $M_p$ for $p \neq r$, which is 
due to the separability of the array.} Inserting $L_r^{\rm opt}$ from \eqref{lopt} and \eqref{loptmod} into expression 
\eqref{msespasmoosingle}, we obtain $\rd{{\rm MSE}}_{{\rm SS}_{\rm opt}} = \rd{{\rm MSE}}_{\rm SS}(L_r^{\rm opt})$ as
\begin{align} 
\rd{{\rm MSE}}_{{\rm SS}_{\rm opt}} \approx \begin{cases} 
\frac{1}{\hat{\rho}} \cdot \frac{27}{4} \cdot \frac{a}{M_r^3} & \text{if } M_r ~ {\rm mod}~3 = 0 \\
\frac{1}{\hat{\rho}} \cdot \frac{27}{4} \cdot \frac{a}{\left( M_r + \frac{1}{2} \right)^2 (M_r - 1)} & \text{if } M_r ~ {\rm mod}~3 = 1 \\ 
\frac{1}{\hat{\rho}} \cdot \frac{27}{4} \cdot \frac{a}{\left( M_r - \frac{1}{2} \right)^2 (M_r + 1)} & \text{if } M_r ~ {\rm mod}~3 = 2, 
\end{cases} 
\label{mse_lopt}
\end{align}
where $a = \prod_{\substack{p = 1 \\ p \neq r}}^R \frac{c_p}{\subp{M}^2 L_p^2}$. It is clear that the MSE 
for a fixed $\hat{\rho}$ is lowest when $M_r$ is a multiple of $3$. Again, for $R=1$, these results are in 
line with those derived in \cite{rao1990spasmoo, rao1993spasmoo,hua1990spasmoo} for harmonic retrieval. 
\subsection{Asymptotic Spatial Smoothing Gain}
Based on the result for $L_r^{\rm opt}$, the maximum asymptotic gain obtained from spatial smoothing can be explicitly 
quantified. To this end, we contrast $\rd{{\rm MSE}}_{\rm SS}(L_r^{\rm opt})$ from above with the result $\rd{{\rm MSE}} = 
\frac{1}{\hat{\rho}} \cdot \frac{M_r}{M(M_r-1)^2}$ from \cite{roemer2012perf} and \cite{steinwandt2014tsp} without spatial 
smoothing. The maximum asymptotic spatial smoothing gain in the $r$-th mode defined as $\rd{\eta}_{\rm SS}(L_r^{\rm opt}) 
= \rd{{\rm MSE}} / \rd{{\rm MSE}}_{\rm SS}(L_r^{\rm opt})$ can be computed as
\begin{equation}
\rd{\eta}_{\rm SS}(L_r^{\rm opt}) \approx 
\begin{cases} 
\frac{4}{27} \cdot \frac{M_r^4}{(M_r - 1)^2} \cdot \frac{1}{M a} & \text{if } M_r ~ {\rm mod}~3 = 0 \\
\frac{4}{27} \cdot \frac{M_r (M_r + \frac{1}{2})^2}{(M_r - 1)} \cdot \frac{1}{M a} & \text{if } M_r ~ {\rm mod}~3 = 1 \\ 
\frac{4}{27} \cdot \frac{M_r (M_r - \frac{1}{2})^2 (M_r + 1)}{(M_r - 1)^2 M a} & \text{if } M_r ~ {\rm mod}~3 = 2. 
\end{cases}
\label{ssgain}
\end{equation}
\subsection{Asymptotic Efficiency of 1-D ESPRIT-type algorithms with Spatial Smoothing}
Furthermore, the optimal value for $L_r^{\rm opt}$ from Subsection \ref{sec:optl} allows to analytically compute the asymptotic 
efficiency of the considered $R$-D ESPRIT-type and $R$-D NC ESPRIT-type algorithms with spatial smoothing for a 
single source. To this end, we utilize the simplified single source expressions of the deterministic $R$-D 
Cram\'{e}r-Rao bound (CRB) and $R$-D NC CRB in \cite{roemer2012perf} and \cite{steinwandt2014tsp}, respectively. 
As both expressions are the same, we here only state the \red{conventional} case from \cite{roemer2012perf}. 

For the case of an $M$-element $R$-D uniform sampling grid with an $M_r$-element ULA in the $r$-th mode and a 
single source ($d=1$), the deterministic $R$-D CRB can be simplified to \cite{roemer2012perf}
\begin{align}
\bm C = \mathrm{diag} \Big\{ \big[ C^{(1)},\ldots,C^{(R)} \big]^\trans \Big\}, 
\label{nccrb_single}
\end{align}
where $C^{(r)} = \frac{1}{\hat{\rho}} \cdot \frac{6}{ M (M^2_r-1)}$. Using \eqref{msespasmoosingle} and 
\eqref{nccrb_single}, the asymptotic efficiency $\rd{\eta}(L_r^{\rm opt}) = \lim_{\hat{\rho} \to \infty} 
\rd{C} / \rd{{\rm MSE}}_{\rm SS}(L_r^{\rm opt})$ of the spatially smoothed versions of $R$-D Standard and $R$-D 
Unitary ESPRIT as well as $R$-D NC Standard and $R$-D NC Unitary ESPRIT can be computed in closed-form for 
arbitrary dimensions $R$. As an example, the asymptotic efficiency $\eta(L^{\rm opt})$ for $R=1$ is given by 
\begin{equation}
\eta(L^{\rm opt}) \approx \begin{cases} 
\frac{8}{9} \cdot \frac{M^2}{M^2-1} & \text{if } M ~ {\rm mod}~3 = 0 \\
\frac{8}{9} \cdot \frac{(M + \frac{1}{2})^2}{M (M+1)} & \text{if } M ~ {\rm mod}~3 = 1 \\ 
\frac{8}{9} \cdot \frac{(M - \frac{1}{2})^2}{M (M-1)} & \text{if } M ~ {\rm mod}~3 = 2. 
\end{cases}
\label{asyeff_single}
\end{equation}
It should be noted that $\eta$ is only a function of the array geometry, i.e., the number of sensors $M$. Moreover, it is straightforward to see that the asymptotic efficiency is larger when $M$ is a multiple of $3$. 
As one of the main results from \eqref{asyeff_single}, we observe that $\lim_{M \to \infty} \eta(L^{\rm opt}) 
= 8/9$ for 1-D \red{ESPRIT-type/NC ESPRIT-type} algorithms with spatial smoothing. In contrast, it was shown in 
\cite{roemer2012perf} and \cite{steinwandt2014tsp} that their counterparts without spatial smoothing 
become less efficient for increasing $M$, i.e., for $M\rightarrow \infty$, we have $\eta \rightarrow 0$. 
Consequently, spatial smoothing provides a significant gain for large $M$.
%
%
%
\section{Simulation Results}
\label{sec:simulations}
In this section, we present two sets of simulation results to assess the behavior of the derived performance 
analysis of ESPRIT-type algorithms based on spatial smoothing and to illustrate the analytical expressions for 
the single source case. 
\subsection{Performance Analysis}
We first compare the square root of the analytical MSE expressions (``ana'') in \eqref{mse}, \eqref{mse_fba}, and 
\eqref{mse_nc} to the root mean square error (RMSE) of the empirical estimation errors (``emp'') of the spatially 
smoothed (SpSm) versions of $R$-D Standard ESPRIT (SE SpSm), $R$-D Unitary ESPRIT (UE SpSm) as well $R$-D NC Standard 
ESPRIT (NC SE SpSm) and $R$-D NC Unitary ESPRIT (NC UE SpSm). For all ESPRIT-type algorithms, LS is used to solve the 
shift invariance equations. We also include the deterministic Cram\'{e}r-Rao bounds for arbitrary signals (Det CRB) 
and strictly SO non-circular sources (Det NC CRB) \cite{steinwandt2015rdnccrb}. 
The RMSE is defined as 
\begin{align}
\mathrm{RMSE} = \sqrt{ \frac{1}{Rd}~\mathbb E\left\{\sum_{r=1}^R \sum_{i=1}^d \left( \rd{\mu}_i - \rd{\hat{\mu}_i} \right)^2 \right\} },
\end{align}
where $\rd{\hat{\mu}_i}$ is the estimate of $i$-th spatial frequency in the $r$-th mode. It is assumed that a known 
number of signals with unit power impinge on uniform array structures consisting of isotropic sensor 
elements with $\lambda/2$-interelement spacing in all dimensions. The phase reference is located at the array 
centroid. The symbols $\bm S_0$ are drawn from a real-valued Gaussian distribution and we assume zero-mean 
circularly symmetric white Gaussian noise. The curves are averaged over 5000 Monte Carlo trials.  

\begin{figure}[t!]
    \centerline{\includegraphics[height=\figheight]{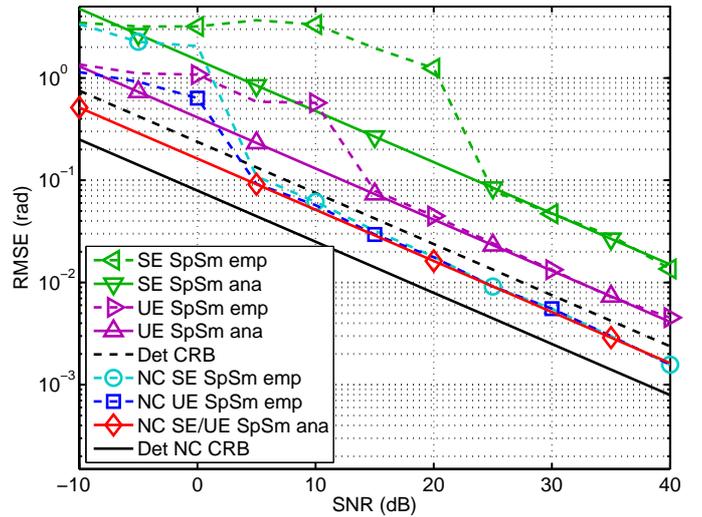}}
    \small
    \caption{RMSE versus SNR for a $6 \times 6 \times 6$ uniform cubic array ($R=3$), and $N=5$, 
    $d=2$ with $\varrho = 0.9$ at $\mu_1^{(1)}=0$, $\mu_2^{(1)}=0.1$, $\mu_1^{(2)}=0$, $\mu_2^{(2)}=0.1$, 
    $\mu_1^{(3)}=0$, $\mu_2^{(3)}=0.1$.} 
    \label{fig:rmse_snr}
    \vspace{-1em}
\end{figure}

\begin{figure}[t!]
    \centerline{\includegraphics[height=\figheight]{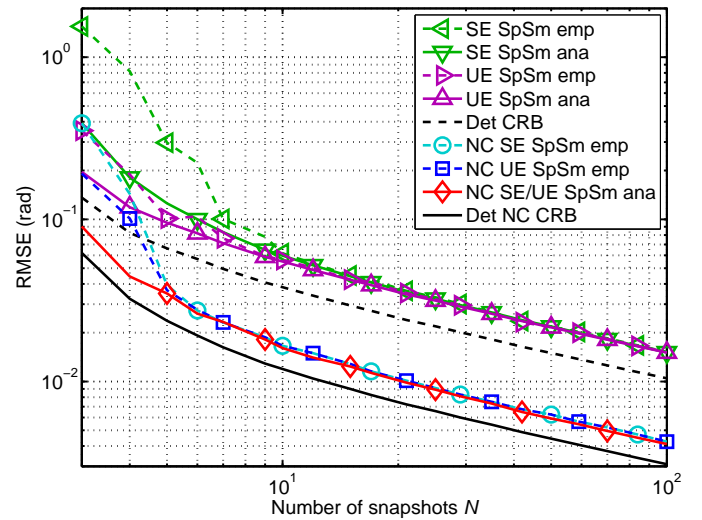}}
    \caption{RMSE versus $N$ for a $6 \times 6$ URA ($R=2$) and SNR $=20$ dB, $d=3$ with $\varrho = 0$  at 
    $\mu_1^{(1)}=0.25$, $\mu_2^{(1)}=0.5$, $\mu_3^{(1)}=0.75$, $\mu_1^{(2)}=0.25$, $\mu_2^{(2)}=0.5$, 
    $\mu_3^{(2)}=0.75$. }
    \label{fig:rmse_snap}
    \vspace{-1em}
\end{figure}

In Fig.~\ref{fig:rmse_snr}, we depict the total RMSE versus the SNR of $d=2$ sources impinging on a $6 \times 6 
\times 6$ uniform cubic array $(R=3)$ with $N=5$. The sources are located at $\mu_1^{(1)} = 0$, $\mu_2^{(1)} = 0.1$, 
$\mu_1^{(2)} = 0$, $\mu_2^{(2)} = 0.1$, $\mu_1^{(3)} = 0$, and $\mu_2^{(3)} = 0.1$. They have a pair-wise correlation 
of $\varrho = 0.9$ and their rotation phases contained in $\bm \Psi$ are given by $\varphi_1 = 0$ and $\varphi_2 = \pi/2$. 
For $L_r$, we choose $L_r^{\rm opt} = M_r/3 = 2$ in each mode, i.e., we have divided the array into a total of $L=8$ subarrays. 
Fig.~\ref{fig:rmse_snap} investigates the total RMSE versus the number of snapshots $N$ for a $6 \times 6$ uniform 
rectangular array (URA) $(R=2)$, where the SNR is 20 dB and $L_r = L_r^{\rm opt} = M_r/3 = 3$. We have $d = 3$ 
uncorrelated ($\varrho = 0$) sources at $\mu_1^{(1)} = 0.25$, $\mu_2^{(1)} = 0.5$, $\mu_3^{(1)} = 0.75$, $\mu_1^{(2)} = 0.25$, $\mu_2^{(2)} 
= 0.5$, and $\mu_3^{(2)} = 0.75$. The rotation phases are given by $\varphi_1 = 0$, $\varphi_2 = \pi/4$, and $\varphi_3 
= \pi/2$. 

It is apparent from Fig.~\ref{fig:rmse_snr} and Fig.~\ref{fig:rmse_snap} that the analytical results agree well with 
the empirical results for high effective SNRs, i.e., either high SNRs or a large sample size. Furthermore, NC SE SpSm 
and NC UE SpSm provide the lowest estimation errors and perform asymptotically identical at high effective SNRs. However, 
NC UE SpSm should be preferred due to its lower complexity and its better performance at low SNRs.
\subsection{Analytical Results for a Single Source}
In this subsection, the derived analytical results (``ana'') in \eqref{mse_lopt} 
and \eqref{asyeff_single} for a single source $(d=1)$ are compared to their empirical versions. We also 
include the analytical and empirical single source results from \cite{roemer2012perf} and \cite{steinwandt2014tsp} 
without spatial smoothing. The source is located at $\rd{\mu} = 0,\forall r$ (however, its location has no impact on 
the MSE) and the effective SNR $\rho$ is 46 dB with $P=1$, $N=4$, and $\sigman^2 = 10^{-4}$.

\begin{figure}[t!]
    \centerline{\includegraphics[height=\figheight]{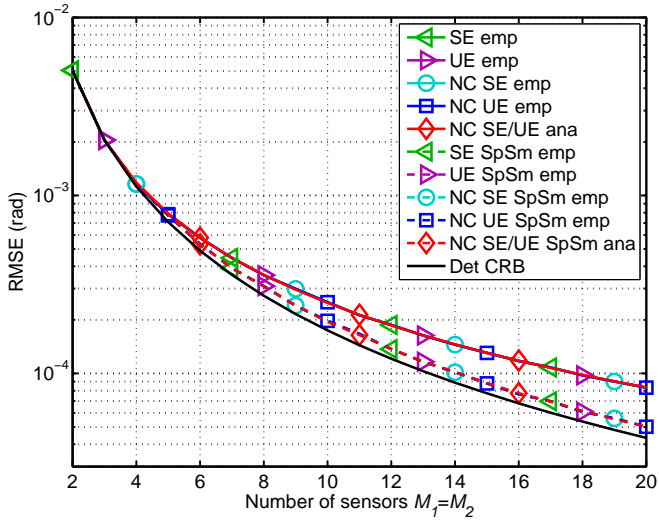}}
    \small
    \caption{RMSE versus $M_1=M_2$ of a $M_1 \times M_2$ URA $(R=2)$ for $d=1$ at $\mu^{(1)} = 0$, $\mu^{(2)} = 0$, 
    and $\rho=46$ dB ($P = 1$, $N = 4$, $\sigma_n^2 = 10^{-4}$).}
    \label{fig:rmse}
    \vspace{-1em}
\end{figure}

Fig.~\ref{fig:rmse} illustrates the total RMSE using \eqref{mse_lopt} as a function of the number of 
sensors $M_1=M_2$ for a 2-D $M_1 \times M_2$ URA. We observe that the spatial smoothing based ESPRIT-type algorithms 
perform considerably closer to the CRB compared to the algorithms without spatial smoothing. 

\begin{figure}[t!]
    \centerline{\includegraphics[height=\figheight]{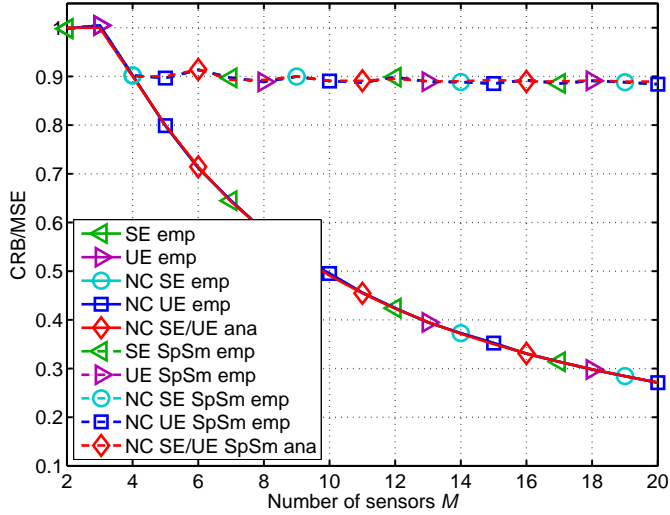}}
    \small
    \caption{Asymptotic efficiency versus $M$ of a ULA $(R=1)$ for $d=1$ at $\mu = 0$ and $\rho = 46$ dB 
    ($P = 1$, $N = 4$, $\sigma_n^2 = 10^{-4}$).}
    \label{fig:eff}
    \vspace{-1em}
\end{figure}

Fig.~\ref{fig:eff} presents the asymptotic efficiency \eqref{asyeff_single} for $R=1$ versus $M$ of a ULA. 
The asymptotic efficiency for the non-spatial smoothing case, i.e., $L=1$, is given by $\eta(L=1) 
= \frac{6(M-1)}{M(M+1)}$. It is clear from Fig.~\ref{fig:eff} that all the algorithms are asymptotically 
efficient for $M=2$ and $M=3$. As $M$ increases further, the efficiency of the algorithms with spatial smoothing 
approaches the value $8/9$, while that of the non-spatial smoothing based algorithms becomes increasingly inefficient. 
Moreover, Fig.~\ref{fig:eff} confirms the observation from \eqref{asyeff_single} that $\eta(L^{\rm opt})$ 
is slightly higher for values of $M$ that are multiples of 3. 
\section{Conclusion}
\label{sec:conclusions}
This paper presents a first-order performance analysis of the spatially smoothed versions of $R$-D Standard 
ESPRIT and $R$-D Unitary ESPRIT for arbitrary sources as well as $R$-D NC Standard ESPRIT and $R$-D NC Unitary 
ESPRIT for strictly SO non-circular sources. 
The derived expressions are asymptotic in the effective SNR and no assumptions on the noise statistics are 
required apart from a zero-mean and finite SO moments. We show that both spatially smoothed $R$-D NC ESPRIT-type 
algorithms perform asymptotically identical in the high effective SNR regime. As the performance generally depends 
on the number of subarrays, we have simplified the derived $R$-D MSE expressions for the special case of a 
single source, which allows to analytically compute the optimal number of subarrays for spatial smoothing. 
Additionally, we have derived the asymptotic spatial smoothing gain and calculated the asymptotic efficiency 
for this special case. The analytical results are supported by simulations. 
\appendices
\section{Proof of Theorem \ref{thm:fba}}
\label{app:fba}
To show this result, we simply use the FBA-processed and spatially smoothed augmented measurement matrix 
%
\begin{align}
\ssmnc{\tilde{\bm X}} & \! = \! \begin{bmatrix} \ssmnc{\bm X} \!\! & \bm \Pi_{2\sub{M}} \ssmncC{\bm X} \bm \Pi_{NL} \end{bmatrix} 
\in\compl^{2\sub{M} \times 2NL}
\label{fba_nc}  
\end{align}
and compute the Gram matrix $\bm G = \ssmnc{\tilde{\bm X}} \ssmncH{\tilde{\bm X}}$, which yields
\begin{align}
\bm G = \ssmnc{\bm X} \ssmncH{\bm X} + \bm \Pi_{2\sub{M}} \ssmncC{\bm X} \ssmncT{\bm X} \bm \Pi_{2\sub{M}}.
\label{gram}
\end{align}
Expanding the second term of \eqref{gram} using \eqref{smoothnc}, we obtain 
\begin{align}
&\bm \Pi_{2\sub{M}} \left( \sum_{\underline{\bm \ell} = \underline{\bm 1}}^{\underline{\bm L}} \nc{\bm J}_{\underline{\bm \ell}} \ncC{\bm X} \ncT{\bm X} 
\ncT{\bm J}_{\underline{\bm \ell}} \right) \bm \Pi_{2\sub{M}} \notag \\
& = \sum_{\underline{\bm \ell} = \underline{\bm 1}}^{\underline{\bm L}} 
	\left[\begin{matrix} 
		\bm \Pi_{\sub{M}} \bm J_{\underline{\bm \ell}} \bm \Pi_{M} \bm X \bm X^\herm \bm \Pi_{M} \bm J_{\underline{\bm \ell}}^\trans \bm \Pi_{\sub{M}}
	\\ \bm \Pi_{\sub{M}} \bm J_{\underline{\bm \ell}} \bm X^\conj \bm X^\herm \bm \Pi_{M} \bm J_{\underline{\bm \ell}}^\trans \bm \Pi_{\sub{M}} \end{matrix} \right. \notag \\ 
&\left. \quad\quad\quad\quad\quad\quad\quad\quad \begin{matrix} 
		\bm \Pi_{\sub{M}} \bm J_{\underline{\bm \ell}} \bm \Pi_{M} \bm X \bm X^\trans \bm J_{\underline{\bm \ell}}^\trans \bm \Pi_{\sub{M}}
	\\ \bm \Pi_{\sub{M}} \bm J_{\underline{\bm \ell}} \bm X^\conj \bm X^\trans \bm J_{\underline{\bm \ell}}^\trans \bm \Pi_{\sub{M}} 
\end{matrix} \right].
\label{gram_mod}
\end{align}
Next, we observe the symmetries $\bm \Pi_{\sub{M}} \bm J_{\underline{\bm \ell}} \bm \Pi_{M} = \bm J_{\underline{\bm L} - \underline{\bm \ell} + \underline{\bm 1}}$ 
and $\bm \Pi_{\sub{M}} \bm J_{\underline{\bm \ell}} = \bm J_{\underline{\bm L} - \underline{\bm \ell} + \underline{\bm 1}} \bm \Pi_{M}$. Hence, we perform 
a change of variables to $\underline{\bm m} = \underline{\bm L} - \underline{\bm \ell} + \underline{\bm 1}$, which simplifies \eqref{gram_mod} to
\begin{align}
& \sum_{\underline{\bm m} = \underline{\bm 1}}^{\underline{\bm L}} 
	\begin{bmatrix} 
		 	 \bm J_{\underline{\bm m}} \bm X \bm X^\herm \bm J_{\underline{\bm m}}^\trans 
		 & \bm J_{\underline{\bm m}} \bm X \bm X^\trans \bm \Pi_{M} \bm J_{\underline{\bm m}}^\trans
		\\ \bm J_{\underline{\bm m}} \bm \Pi_{M} \bm X^\conj \bm X^\herm \bm J_{\underline{\bm m}}^\trans 
		 & \bm J_{\underline{\bm m}} \bm \Pi_{M} \bm X^\conj \bm X^\trans \bm \Pi_{M} \bm J_{\underline{\bm m}}^\trans 
	\end{bmatrix}  \notag \\
& = \ssmnc{\bm X} \ssmncH{\bm X}.
\label{gram_mod2}
\end{align} 
Replacing the second term of \eqref{gram} by \eqref{gram_mod2}, we have $\bm G = 2 \cdot \ssmnc{\bm X} \ssmncH{\bm X}$. 
\red{Thus, the matrix $\bm G$ reduces to the scaled Gram matrix of $\ssmnc{\bm X}$, i.e.,} the column space of $\ssmnc{\bm X}$ is the 
same as the column space of the Gram matrix of $\ssmnc{\bm X}$. Consequently, FBA has no effect on the column space of $\ssmnc{\bm X}$. This 
completes the proof. \qed
\section{Proof of Theorem \ref{thm:ncunit_single}} 
\label{app:ncunit_single}
This theorem consists of several parts, which we address in separate subsections. 
\subsection{MSE for $R$-D Standard ESPRIT with Spatial Smoothing}
We start the proof by simplifying the MSE expression for $R$-D Standard ESPRIT with spatial smoothing 
in \eqref{mse} and for $d=1$. In the single source case the noise-free spatially smoothed measurement matrix $\ssmnought{\bm X} 
\in\compl^{\sub{M} \times NL}$ can be written as
\begin{align}
\ssmnought{\bm X} & = \ssm{\bar{\bm a}}(\bm \mu) \bm \phi^\trans \left( \bm I_L \otimes \bar{\bm s}^\trans \right) 
= \ssm{\bar{\bm a}}(\bm \mu) \bm a_L^\trans  \left( \bm I_L \otimes \bar{\bm s}^\trans \right)  \notag\\
& = \ssm{\bar{\bm a}}(\bm \mu) \left( \bm a_L  \otimes \bar{\bm s} \right)^\trans = \ssm{\bar{\bm a}}(\bm \mu) \bar{\bm s}_L^\trans
\label{modsingle}
\end{align}
where $\ssm{\bar{\bm a}}(\bm \mu) = \bar{\bm a}_1^{(1)}(\mu^{(1)}) \otimes \cdots \otimes \bar{\bm a}_1^{(R)}(\mu^{(R)}) 
\in\compl^{\sub{M} \times 1}$ is the spatially smoothed array steering vector in all $R$ modes with 
$\rd{\bar{\bm a}}_1(\rd{\mu}) = \bm J^{(M_r)}_{1_r} \rd{\bar{\bm a}}(\rd{\mu}) \in\compl^{\subr{M} \times 1},~r=1,\ldots,R$ 
and $\bm \phi = \bm a_L = \bm a_{L_1}^{(1)}(\mu^{(1)}) \otimes \cdots \otimes \bm a_{L_R}^{(R)}(\mu^{(R)}) 
\in\compl^{L \times 1}$ with $\rd{\bm a}_{L_r}(\rd{\mu}) = [1,\e^{\j\rd{\mu}},\ldots,\e^{\j\rd{\mu} (L_r-1)}]^\trans 
\in\compl^{L_r \times 1},~\forall~r$. Moreover, $\bar{\bm s} \in \compl^{N \times 1}$ contains the source symbols 
with the empirical source power $\hat{P}_{\rm s} = \twonorm{\bar{\bm s}}^2 / N$ and we have $\bar{\bm s}_L^\herm 
\bar{\bm s}_L = N L \hat{P}_{\rm s}$. In what follows, we drop the dependence of $\ssm{\bar{\bm a}}(\bm \mu)$ on $\bm \mu$ 
for notational convenience. For a ULA of isotropic elements in each of the $R$ modes, $\rd{\bar{\bm a}}$ is 
given by \eqref{asteer_phasecen} and $\twonorm{\ssm{\bar{\bm a}}}^2 = \sub{M} 
= M-L+1$. The selection matrices $\ssmone{\rdtil{\bm J}}$ and $\ssmtwo{\rdtil{\bm J}}$ are chosen as
$\ssmone{\rdtil{\bm J}} = [\bm I_{\subr{M} - 1},\bm 0_{(\subr{M} - 1)\times 1}]$ and $\ssmtwo{\rdtil{\bm J}} 
= [\bm 0_{(\subr{M} - 1)\times 1},\bm I_{\subr{M}-1}]$ for maximum overlap, i.e., $\subsel{\subr{M}} = 
\subr{M}-1$. Note that \eqref{modsingle} is a rank-one matrix and we can directly determine the subspaces 
from the SVD as
\begin{align}
\ssms{\bm U} & = \ssms{\bm u} = \frac{\ssm{\bar{\bm a}}}{\twonorm{\ssm{\bar{\bm a}}}} 
= \frac{\ssm{\bar{\bm a}}}{\sqrt{\sub{M}}} \notag\\
\ssms{\bm \Sigma} & = \ssms{\sigma} = \sqrt{\sub{M} N L\hat{P}_{\rm s}} \notag\\
\ssms{\bm V} & = \ssms{\bm v} = \frac{\bar{\bm s}_L^\conj}{\twonorm{\bar{\bm s}_L}} 
= \frac{\bar{\bm s}_L^\conj}{\sqrt{N L \hat{P}_{\rm s}}}. \notag
\end{align}
For the MSE expression in \eqref{mse}, we also require $\bm P^{\perp}_{\ssm{\bar{\bm a}}} = \ssmn{\bm U^{}} 
\ssmnH{\bm U} = \bm I_{\sub{M}} - \frac{1}{\sub{M}} \ssm{\bar{\bm a}}^{} \ssm{\bar{\bm a}}^\herm$, which is the 
projection matrix onto the noise subspace. Moreover, we have $\rd{\bm \Phi} = \expof{\j \rd{\mu}}$ and 
hence, the eigenvectors are $\rd{\bm p_i} = \rd{\bm q_i} = 1$. The SO moments $\ssm{\bm R}$ 
and $\ssm{\bm C}$ of the noise are given by \eqref{covss} with $\bm R_{\rm nn} = 
\sigman^2 \bm I_M$ and $\bm C_{\rm nn} = \bm 0$.  
 
Inserting these expressions into \eqref{mse}, we get 
\begin{align}
\expvof{(\Delta \rd{\mu})^2} = \frac{1}{2} \cdot \bm z^\herm \ssm{\bm R}^\trans \bm z = \frac{1}{2} \cdot \bm z^\trans \ssm{\bm R} \bm z^\conj 
\label{mse_single}
\end{align}
with $\bm z = \ssm{\bm W}^\trans {\rd{\ssmi{\bm r}}}$ and
\small
\begin{align}
{\rd{\ssmi{\bm r}}} & = \left[ \left(\ssmone{\rdtil{\bm J}}  \frac{\ssm{\bar{\bm a}}}{\sqrt{\sub{M}}} \right)^+ \!\!
\left(\ssmtwo{\rdtil{\bm J}} /\expof{\j \rd{\mu}} - \ssmone{\rdtil{\bm J}} \right)\right]^\trans \in\compl^{\sub{M} \times 1} , \notag \\
\ssm{\bm W} & = \left(\frac{1}{\sqrt{\sub{M} N L \hat{P}_{\rm s}}} \cdot \frac{\bar{\bm s}_L^\herm}{\sqrt{N L \hat{P}_{\rm s}}}\right) 
\otimes \bm P^{\perp}_{\ssm{\bar{\bm a}}} \in\compl^{\sub{M} \times \sub{M} N L}. \notag 
\end{align}
\normalsize
Note that the term $\bm z^\trans$ can also be written as $\bm z^\trans = \tilde{\bm s}^\trans \otimes 
\rdT{\tilde{\bm a}}$, where

\vspace{-1em}
\small
\begin{align}
\tilde{\bm s}^\trans &= \frac{1}{\sqrt{\sub{M} N L \hat{P}_{\rm s}}} \cdot \frac{(\bm a_L \otimes \bar{\bm s})^\herm}
{\sqrt{N L \hat{P}_{\rm s}}} \\
\rdT{\tilde{\bm a}} & =  \left(\ssmone{\rdtil{\bm J}} \frac{\ssm{\bar{\bm a}}}{\sqrt{\sub{M}}} \right)^+\!\!\!\left(\ssmtwo{\rdtil{\bm J}} /
\expof{\j\rd{\mu}} - \ssmone{\rdtil{\bm J}} \right) \bm P^{\perp}_{\ssm{\bar{\bm a}}}. 
\label{a_tilde}
\end{align}
\normalsize
Next, we further simplify the expression $\rdT{\tilde{\bm a}}$ and expand the pseudo-inverse of 
$\ssmone{\rdtil{\bm J}} \ssm{\bar{\bm a}}$ using the relation $\bm x^\pinv = \bm x^\herm / \twonorm{\bm x}^2$. 
As $\ssmone{\rdtil{\bm J}}$ selects $\subr{M} - 1$ out of $\subr{M}$ elements in the $r$-th mode, we have $\big\| 
\ssmone{\rdtil{\bm J}} \ssm{\bar{\bm a}} \big\|_2^2 = \frac{\sub{M}}{\subr{M}} \cdot (\subr{M} - 1)$. Then, taking 
the shift invariance equation $\ssmtwo{\rdtil{\bm J}} \ssm{\bar{\bm a}} / \expof{\j\rd{\mu}} - \ssmone{\rdtil{\bm J}} 
\ssm{\bar{\bm a}} = \bm 0$ in the $r$-th mode into account, we obtain
\begin{align}
\rdT{\tilde{\bm a}} & = \frac{\sqrt{\sub{M}} \subr{M}}{\sub{M} (\subr{M}-1)} \cdot \rdT{\check{\bm a}},  \\
\rdT{\check{\bm a}} & = \ssm{\bar{\bm a}}^\herm \left( \ssmtwo{\rdtilH{\bm J}} \ssmtwo{\rdtil{\bm J}}  - 
\ssmone{\rdtilH{\bm J}} \ssmone{\rdtil{\bm J}} \right).
\label{asingle}
\end{align}
Since the vector $\ssm{\bar{\bm a}}$ and the matrices $\ssmk{\rdtil{\bm J}},~k=1,2$, contained in $\rdT{\bar{\bm a}}$ can 
be written as $\ssm{\bar{\bm a}} = \bar{\bm a}_1^{(1)} \otimes \cdots \otimes \bar{\bm a}_1^{(R)}$ and $\ssmk{\rdtil{\bm J}} = 
\bm I_{\prod_{l=1}^{r-1} \subl{M}} \otimes \rd{\ssmk{\bm J}} \otimes \bm I_{\prod_{l=r+1}^{R} \subl{M}}$, all the 
unaffected modes can be factored out of $\rdT{\check{\bm a}}$, yielding 
\begin{align}
\rdT{\check{\bm a}} & = \left( \bar{\bm a}_1^{(1)} \otimes \cdots \otimes \bar{\bm a}_1^{(r-1)} \right)^\herm \otimes \rdT{\breve{\bm a}}_1  \notag \\
& \quad \otimes \left(\bar{\bm a}_1^{(r+1)} \otimes \cdots \otimes \bar{\bm a}_1^{(R)} \right)^\herm,
\label{asingle2}
\end{align}
where we have $\rdT{\breve{\bm a}}_1  = \rdH{\bar{\bm a}}_1 \Big( \rdH{\ssmtwo{\bm J}} \rd{\ssmtwo{\bm J}} - 
\rdH{\ssmone{\bm J}} \rd{\ssmone{\bm J}} \Big)$ with 
$\rdH{\bar{\bm a}}_1 \!\!\!\!= \big[ \expof{\j \frac{(M_r-1)}{2}\rd{\mu}}\!\!\!, 
\ldots, \expof{-\j \frac{(M_r-2L_r-1)}{2} \rd{\mu}}\!\!\!, \expof{-\j \frac{(M_r-2L_r+1)}{2} \rd{\mu}} \big]$. 
%
%
Then, 
it is easy to verify that
\begin{align}
\rdT{\breve{\bm a}}_1 & = \left[ -\expof{\j \frac{(M_r-1)}{2}\rd{\mu}}, 0, \ldots, 0, \expof{-\j \frac{(M_r-2L_r+1)}{2} \rd{\mu}} \right].
\notag
\end{align}
Thus, the MSE expression in \eqref{mse_single} is given by
\begin{align} 
&\expvof{(\Delta \rd{\mu})^2} = \frac{k^2}{2} \cdot  \bm v^\trans \ssm{\bm R} \bm v^\conj,
\label{msesingle2}
\end{align}
where we haved used $\bm z^\trans = k \cdot \bm v^\trans$ with $\bm v^\trans = \bm a_L^\herm \otimes \bm s^\herm \otimes \rdT{\check{\bm a}}$ and
%
$k = \frac{1}{N L \hat{P}_{\rm s}} \cdot \frac{ \subr{M}}{ \sub{M} (\subr{M}-1)}$.
%
After straightforward calculations, we further simplify \eqref{msesingle2} to obtain \eqref{msesingle_temp}-\eqref{msesingle3} 
at the bottom of this page,
%
%
%
\begin{figure*}[bp]
\hrule 
\begin{align} 
&\expvof{(\Delta \rd{\mu})^2} = \frac{k^2}{2} \cdot \sigman^2 \cdot \bm s^\herm \bm s \cdot \sum_{\underline{\bm \ell} = \underline{\bm 1}}^{\underline{\bm L}} \sum_{\underline{\bm m} = \underline{\bm 1}}^{\underline{\bm L}} \left( 
\left( \prod_{\substack{p = 1 \\ p \neq r}}^R \bm a_1^{(p)^\herm} \bm J_{\ell_p}^{(M_p)} \bm J_{m_p}^{{(M_p)}^\trans} \bm a_1^{(p)} \right) 
\cdot \rdT{\breve{\bm a}}_1  \bm J_{\ell_r}^{(M_r)} \bm J_{m_r}^{{(M_r)}^\trans} \rdC{\breve{\bm a}}_1  
\cdot \e^{\j \sum_{s=1}^R \mu^{(s)} (\ell_s - m_s) } \right) \label{msesingle_temp} \\
& = \frac{k^2}{2} \cdot \sigman^2 \cdot N \hat{P}_{\rm s} \cdot \left( \prod_{\substack{p = 1 \\ p \neq r}}^R \bm a_1^{(p)^\herm} \left( \sum_{\ell_p = 1}^{L_p} \sum_{m_p = 1}^{L_p} \bm J_{\ell_p}^{(M_p)} \bm J_{m_p}^{{(M_p)}^\trans} \right) \bm a_1^{(p)} \right) \cdot \rdT{\breve{\bm a}}_1 \!\! 
\left( \sum_{\ell_r = 1}^{L_r} \sum_{m_r = 1}^{L_r} \bm J_{\ell_r}^{(M_r)} \bm J_{m_r}^{{(M_r)}^\trans} \!\! \cdot \e^{\j \sum_{s=1}^R \mu^{(s)} (\ell_s - m_s) }\right) \rdC{\breve{\bm a}}_1  \label{msesingle3_tmp} \\
& = \frac{k^2}{2} \cdot \sigman^2 \cdot N \hat{P}_{\rm s} \cdot \left( \prod_{\substack{p = 1 \\ p \neq r}}^R c_p \right) \cdot 2 \cdot \min\{L_r, M_r - L_r\}
\label{msesingle3}
\end{align} 
\end{figure*}
where $c_p$ in \eqref{msesingle3} is given by \eqref{cp} and
%
%
it can be shown that the last term in \eqref{msesingle3_tmp} evaluates to $2\cdot(L_r - \max\{2 \cdot L_r - M_r,0\}) = 2 \cdot \min\{L_r, M_r - L_r\}$. 
Consequently, the MSE of $R$-D Standard ESPRIT with spatial smoothing is given by
\begin{align} 
& \expvof{(\Delta \rd{\mu})^2} = \frac{\sigman^2}{N \hat{P}_{\rm s}} \cdot  \frac{ \subr{M}^2 \min\{L_r,M_r - L_r\}}{L^2 \sub{M}^2 (\subr{M}-1)^2} \cdot \prod_{\substack{p = 1 \\ p \neq r}}^R c_p \notag \\[-1ex]
& = \frac{\sigman^2}{N \hat{P}_{\rm s}} \cdot  \frac{\min\{L_r,M_r - L_r\}}{(M_r - L_r)^2 L_r^2 } \cdot 
\prod_{\substack{p = 1 \\ p \neq r}}^R \frac{c_p}{\subp{M}^2 L_p^2},
\label{msesingle4}
\end{align}
where we have used the fact that $\sub{M} = \subr{M} \cdot \prod_{\substack{p = 1 \\ p \neq r}}^R \subp{M}$ and 
$L = L_r \cdot \prod_{\substack{p = 1 \\ p \neq r}}^R L_p$. Equation \eqref{msesingle4} is the desired result. \qed
\subsection{MSE for $R$-D Unitary ESPRIT with Spatial Smoothing}
The second part of the theorem is to show that for a single source, the MSE of $R$-D Unitary ESPRIT with 
spatial smoothing in \eqref{mse_fba} is the same as the MSE of $R$-D Standard ESPRIT with spatial smoothing 
in \eqref{mse}. Firstly, we simplify $\ssmnought{\tilde{\bm X}}$ from \eqref{fba} and find
\begin{align}
\ssmnought{{\tilde{\bm X}}} & = \begin{bmatrix} \ssm{\bar{\bm a}} \bar{\bm s}_L^\trans & 
\bm \Pi_{\sub{M}} \ssm{\bar{\bm a}}^\conj \bar{\bm s}_L^\herm \bm \Pi_{NL} \end{bmatrix} \label{mod1}\\
& = \ssm{\bar{\bm a}} \begin{bmatrix} \bar{\bm s}_L^\trans & \e^{\j\sum_{r=1}^R (L_r - 1) \rd{\mu}} \bar{\bm s}_L^\herm \bm \Pi_{NL} \end{bmatrix} \notag \\
& = \ssm{\bar{\bm a}} \bar{\bar{\bm s}}_L^\trans,
\label{modsinglefba}
\end{align}
where in \eqref{mod1}, we have used the fact that $\bm \Pi_{\subr{M}} \rdC{\bar{\bm a}}_1(\rd{\mu}) = 
\rd{\bar{\bm a}}_1(\rd{\mu}) \e^{\j (L_r - 1) \rd{\mu}}$ holds for a ULA in the $r$-th mode. Moreover, we have defined 
\begin{align}
\bar{\bar{\bm s}}_L = \begin{bmatrix} \bar{\bm s}_L \\ \e^{\j\sum_{r=1}^R (L_r - 1) \rd{\mu}} \bm \Pi_{NL} \bar{\bm s}_L^\conj \end{bmatrix} 
= \begin{bmatrix} \bm a_L \otimes \bar{\bm s} \\ \bm a_L \otimes \bm \Pi_N \bar{\bm s}^\conj \end{bmatrix}.
\end{align}
Note that $\twonorm{\bar{\bar{\bm s}}_L}^2 = 2NL \hat{P}_{\rm s}$. The subspaces from the SVD of 
$\ssmnought{{\tilde{\bm X}}}$ are obtained as
\begin{align}
\ssms{\tilde{\bm u}} & = \frac{\ssm{\bar{\bm a}}}{\sqrt{\sub{M}}} = \ssms{\bm u}, \quad
\ssms{\tilde{\sigma}} = \sqrt{2\sub{M} N L\hat{P}_{\rm s}} \notag\\
\ssms{\tilde{\bm v}} & = \frac{\bar{\bar{\bm s}}_L^\conj}{\sqrt{2N L \hat{P}_{\rm s}}}. \notag
\end{align}
Compared to the previous subsection, it is apparent that FBA does not affect the column space $\ssms{\bm u}$, 
such that $\ssmn{\tilde{\bm U^{}}} = \ssmn{\bm U^{}}$ and thus $\tilde{\bm P}^{\perp}_{\ssm{\bar{\bm a}}} = 
\bm P^{\perp}_{\ssm{\bar{\bm a}}}$. However, FBA destroys the circular symmetry of the noise, resulting 
in an additional term in the MSE expression. Following the derivation for $R$-D Standard ESPRIT 
with spatial smoothing, it can be shown that $\tilde{\bm z}^\trans = {\rdT{\ssmi{\tilde{\bm r}}}} \ssm{\tilde{\bm W}} = 
\tilde{\tilde{\bm s}}^\trans \otimes \rdT{\tilde{\bm a}}$, where
\begin{align}
\tilde{\tilde{\bm s}}^\trans & = \frac{1}{\sqrt{2 \sub{M} N L \hat{P}_{\rm s}}} \cdot \frac{\bar{\bar{\bm s}}_L^\herm}
{\sqrt{2 N L \hat{P}_{\rm s}}}
\end{align}
and $\rdT{\tilde{\bm a}}$ is given as in \eqref{a_tilde}. Thus, the MSE expression for 
$R$-D Unitary ESPRIT with spatial smoothing in \eqref{mse_fba} can be written as
\begin{align} 
&\expvof{(\Delta \rd{\mu})^2} = \frac{1}{2} \cdot \left( \tilde{\bm z}^\trans 
\ssmtil{\bm R} \tilde{\bm z}^\conj - \mathrm{Re}\left\{ \tilde{\bm z}^\trans \ssmtil{\bm C} 
\tilde{\bm z} \right\} \right) 
\label{msesingle_fba}
\end{align}
where $\ssmtil{\bm R} = (\bm I_2 \otimes \ssm{\bm R})$ and $\ssmtil{\bm C} = (\bm \Pi_2 
\otimes \bm \Pi_{\sub{M}NL} \ssm{\bm R})$. Expanding \eqref{msesingle_fba}, we have

\vspace{-1em}
\small
\begin{align} 
&\expvof{(\Delta \rd{\mu})^2} = \frac{\tilde{k}^2}{2} \cdot \left( \bm v^\trans \ssm{\bm R} \bm v^\conj  
+ \bar{\bm v}^\trans \ssm{\bm R} \bar{\bm v}^\conj \right. \notag  \\
& \left. - \mathrm{Re} \left\{ \bm v^\trans \bm \Pi_{\sub{M}NL} \ssm{\bm R} \bar{\bm v} 
+ \bar{\bm v}^\trans \bm \Pi_{\sub{M}NL} \ssm{\bm R} \bm v^\conj  \right\} \right) 
\label{msesingle_fba2}
\end{align}
\normalsize
where $\tilde{\bm z}^\trans = \tilde{k} \cdot \tilde{\bm v}^\trans$ with $\tilde{\bm v}^\trans = [\bm v^\trans, 
\bar{\bm v}^\trans]$, $\bar{\bm v}^\trans = \bm a_L^\herm \otimes \bm s^\trans \bm \Pi_N \otimes \rdT{\check{\bm a}}$, 
and $\tilde{k} = \frac{1}{2 N L \hat{P}_{\rm s}} \cdot \frac{ \subr{M}}{ \sub{M} (\subr{M}-1)}$. 
Note that the first term of \eqref{msesingle_fba2} was already computed in \eqref{msesingle3} as $2 \cdot \sigman^2 
\cdot N \hat{P}_{\rm s} \cdot \prod_{\substack{p = 1 \\ p \neq r}}^R c_p \cdot \min\{L_r, M_r - L_r\}$. 
The remaining terms can be computed accordingly, where for the second term, we also obtain $2 \cdot \sigman^2 
\cdot N \hat{P}_{\rm s} \cdot \prod_{\substack{p = 1 \\ p \neq r}}^R c_p \cdot \min\{L_r, M_r - L_r\}$ 
while the third and fourth terms both evaluate to $-2 \cdot \sigman^2 \cdot N \hat{P}_{\rm s} \cdot 
\prod_{\substack{p = 1 \\ p \neq r}}^R c_p \cdot \min\{L_r, M_r - L_r\}$.
Inserting these intermediate results into \eqref{msesingle_fba2}, we obtain for the MSE of $R$-D Unitary 
ESPRIT with spatial smoothing

\vspace{-1em}
\small
\begin{align} 
& \expvof{(\Delta \rd{\mu})^2} = \frac{\sigman^2}{N \hat{P}_{\rm s}} \! \cdot \!  \frac{\min\{L_r,M_r - L_r\}}{(M_r - L_r)^2 L_r^2 } \! \cdot \! 
\prod_{\substack{p = 1 \\ p \neq r}}^R \frac{c_p}{\subp{M}^2 L_p^2},
\label{msesingle_fba4}
\end{align}
\normalsize
which is equal to \eqref{msesingle4} and hence proves this part.
\subsection{MSE for $R$-D NC Standard ESPRIT and Unitary ESPRIT with Spatial Smoothing}
The third part of the theorem is to show that the MSE of the spatially smoothed versions of $R$-D NC Standard 
ESPRIT and $R$-D NC Unitary ESPRIT is the same as the MSE for $R$-D Standard ESPRIT and Unitary ESPRIT. 
As we have already proven that the performance of $R$-D NC Standard and $R$-D NC Unitary ESPRIT with spatial 
smoothing is identical in the high effective SNR in general, this must also hold true for the case $d=1$. Hence, 
it is sufficient to simplify the MSE of $R$-D NC Standard ESPRIT in \eqref{mse_nc} for this special case. 

We start by writing $\nc{\ssmnought{\bm X}}$ in \eqref{smoothnc2} as
\begin{align}
\ssmnought{\bm X} & = \nc{\ssm{\bar{\bm a}}} \bar{\bm s}_L^\trans,
\label{modsinglenc}
\end{align}
where $\bar{\bm s}_L$ was defined in \eqref{modsingle} and $\nc{\ssm{\bar{\bm a}}} = [1,\tilde{\Psi}]^\trans \otimes 
\ssm{\bar{\bm a}}$ with $\tilde{\Psi} = \Psi^\conj \Psi^\conj = \e^{-\j2\varphi}$. This follows from \eqref{smoothnc2} 
and the fact that $\nc{\bar{\bm a}}= [1,\tilde{\Psi}]^\trans \otimes \bar{\bm a}$ for a uniform $R$-D array whose phase 
reference is at the centroid, i.e. $\bm \Pi_M \bar{\bm a}^\conj = \bar{\bm a}$ holds. Therefore, we have 
$\big\|\nc{\ssm{\bar{\bm a}}}\big\|_2^2 = 2\sub{M}$. The selection matrices $\ssmk{\ncrtil{\bm J}},~k=1,2$ are given 
by $\ssmk{\ncrtil{\bm J}} = \bm I_2 \otimes \ssmk{\rdtil{\bm J}}$. The SVD of \eqref{modsinglenc} can be explicitly 
expressed as
\begin{align}
\ssms{\nc{\bm u}} & = \frac{\nc{\ssm{\bar{\bm a}}}}{\sqrt{2\sub{M}}}, \quad
\ssms{\nc{\sigma}} = \sqrt{2\sub{M} N L\hat{P}_{\rm s}} \notag\\
\ssms{\nc{\bm v}} & = \frac{\bar{\bm s}_L^\conj}{\sqrt{N L \hat{P}_{\rm s}}} = \ssms{\bm v}. \notag
\end{align}
It is evident that the NC preprocessing only affects the column space $\ssms{\nc{\bm u}}$ while the row space 
$\ssms{\bm v}$ of $R$-D Standard ESPRIT remains unaffected. Therefore, we have $\bm P^{\perp}_{\ssmnc{\bar{\bm a}}} 
= \ssmncn{\bm U} \ssmncnH{\bm U} = \bm I_{\sub{M}} - \frac{1}{\sub{M}} \ssmnc{\bar{\bm a}} \ssmncH{\bar{\bm a}}$. Similarly 
to FBA, the circular symmetry of the noise is destroyed by the NC preprocessing step. In the NC case, it 
can be shown that $\ncT{\bm z} = {\ncrT{\ssmi{\bm r}}} \ssmnc{\bm W} = \ncT{\tilde{\bm s}} \otimes \ncrT{\tilde{\bm a}}$, 
where 

\vspace{-1em}
\small
\begin{align}
\ncT{\tilde{\bm s}} &= \frac{1}{\sqrt{2\sub{M} N L \hat{P}_{\rm s}}} \cdot \frac{(\bm a_L \otimes \bar{\bm s})^\herm}
{\sqrt{N L \hat{P}_{\rm s}}} \label{s_nc}\\
\ncrT{\tilde{\bm a}} \!\!\!& = \!\!  \left(\ssmone{\ncrtil{\bm J}} \frac{\ssmnc{\bar{\bm a}}}{\sqrt{2\sub{M}}} \right)^+\!\!\!\!\left(\ssmtwo{\ncrtil{\bm J}} /
\expof{\j\rd{\mu}} \!\!\!\!\! - \ssmone{\ncrtil{\bm J}} \right) \bm P^{\perp}_{\ssmnc{\bar{\bm a}}}. 
\label{a_tildenc}
\end{align}
\normalsize
Following the lines of the derivation of $R$-D Standard ESPRIT with spatial smoothing, $\ncrT{\tilde{\bm a}}$ 
can be simplified as 
\begin{align}
\ncrT{\tilde{\bm a}} 
& = \frac{\sqrt{2\sub{M}} \subr{M}}{2\sub{M} (\subr{M}-1)} \cdot \begin{bmatrix} 1 \\ \tilde{\Psi} \end{bmatrix}^\herm 
\otimes \rdT{\check{\bm a}},
\label{asinglenc}
\end{align}
where $\rdT{\bar{\bm a}}$ is given in \eqref{asingle2}. Consequently, the MSE for $R$-D NC Standard ESPRIT with spatial 
smoothing in \eqref{mse_nc} can be written as

\vspace{-1em}
\small
\begin{align} 
&\expvof{(\Delta \rd{\mu})^2} = \frac{1}{2} \cdot \left( \ncT{\bm z} 
\ssmnc{\bm R} \ncC{\bm z} \!\!- \mathrm{Re}\left\{ \ncT{\bm z} \ssmnc{\bm C} \bm z \right\} \right), 
\label{msesingle_fbanc}
\end{align}
\normalsize
where $\ssmnc{\bm R}$ and $\ssmnc{\bm C}$ are given according to \eqref{covssnc}.
Next, we use \eqref{s_nc} and \eqref{asinglenc} to express \eqref{msesingle_fbanc} as 

\vspace{-1em}
\small
\begin{align} 
\expvof{(\Delta \rd{\mu})^2} & = \frac{{\nc{k}}^2}{2} \cdot \left( \ncT{\bm v} \ssmnc{\bm R} \ncC{\bm v} \right. \notag \\
& \quad\quad\quad\quad\quad\quad \left. - \mathrm{Re}\left\{ \ncT{\bm v} \ssmnc{\bm C} \nc{\bm v}  \right\} \right), 
\label{msesingle_fba2nc}
\end{align}
\normalsize
where again $\ncT{\bm z} = \nc{k} \cdot \ncT{\bm v}$ with $\ncT{\bm v} = \bm a_L^\herm \otimes \bar{\bm s}^\herm 
\otimes \begin{bmatrix} 1 \\ \tilde{\Psi} \end{bmatrix}^\herm \otimes \rdT{\check{\bm a}}$ and $\nc{k} = 
\frac{1}{2 N L \hat{P}_{\rm s}} \cdot \frac{ \subr{M}}{ \sub{M} (\subr{M}-1)}$.
Considering the first term of \eqref{msesingle_fba2nc} and expanding $\ssmnc{\bm R}$, we apply the same steps 
as in \eqref{msesingle_temp} and \eqref{msesingle3}. As a result, the first term reduces to $4 \cdot \sigman^2 
\cdot N \hat{P}_{\rm s} \cdot \prod_{\substack{p = 1 \\ p \neq r}}^R c_p \cdot \min\{L_r, M_r - L_r\}$. 
The second term of \eqref{msesingle_fba2nc} can be computed accordingly to obtain $- 4 \cdot \sigman^2 
\cdot N \hat{P}_{\rm s} \cdot \prod_{\substack{p = 1 \\ p \neq r}}^R c_p \cdot \min\{L_r, M_r - L_r\}$. 

Using these expressions in \eqref{msesingle_fba2nc}, the MSE of $R$-D NC Standard ESPRIT with spatial 
smoothing is given by

\vspace{-1em}
\small
\begin{align} 
& \expvof{(\Delta \rd{\mu})^2} = \frac{\sigman^2}{N \hat{P}_{\rm s}} \! \cdot \!  \frac{\min\{L_r,M_r - L_r\}}{(M_r - L_r)^2 L_r^2 } \! \cdot \!
\prod_{\substack{p = 1 \\ p \neq r}}^R \frac{c_p}{\subp{M}^2 L_p^2}.
\label{msesingle_fba4nc}
\end{align}
\normalsize
As this result is equal to \eqref{msesingle4} and \eqref{msesingle_fba4}, the theorem has been proven. \qed
\red{
\section{Proof of Equation \eqref{lopt}} 
\label{app:lopt}
For the proof, we consider the case $L_r \leq \frac{M_r}{2}$, however, the derivation for $L_r > \frac{M_r}{2}$ 
follows the same steps. The MSE in \eqref{msespasmoosingle} is given by
\begin{align} 
\rd{{\rm MSE}}_{\rm SS} \approx \frac{1}{\hat{\rho}} \cdot \frac{a}{(M_r - L_r)^2 L_r} 
~\quad \text{for } \quad L_r \leq \frac{M_r}{2},
\label{msespasmoosinglee}
\end{align}
where we have defined $a = \prod_{\substack{p = 1 \\ p \neq r}}^R \frac{c_p}{\subp{M}^2 L_p^2}$. In order to 
determine the optimal number of subarrays $L_r$ in the $r$-th mode, we minimize \eqref{msespasmoosingle} with 
respect to $L_r$. That is, we first compute the derivative of \eqref{msespasmoosingle} with respect to $L_r$ 
given by 
\begin{align} 
\frac{\partial \rd{{\rm MSE}}_{\rm SS}}{\partial L_r} = \frac{1}{\hat{\rho}} \cdot \frac{a(M_r -3 L_r)}{(L_r - M_r)^3 L_r^2}
\label{diff_msespasmoosingle}
\end{align}
and then equate \eqref{diff_msespasmoosingle} to zero and solve for $L_r$, yielding
\begin{align} 
L_r^{\rm opt} = \frac{1}{3} \cdot M_r,
\label{l_r_opt}
\end{align}
which is the desired result in \eqref{lopt}. \qed }
%
%
%
\bibliographystyle{IEEEbib}
\bibliography{refs_perf_spatial_smoothing}

\end{document}